\documentclass[10pt]{IEEEtran}
\usepackage{amsmath}
\usepackage{amsthm}
\usepackage{amsfonts}
\usepackage{amssymb}
\usepackage{makeidx}
\usepackage{cite}
\usepackage{graphicx,bigints}
\usepackage{mathtools}
\usepackage{cases}
\usepackage{color}
\usepackage{hyperref}
\usepackage{bbm}
\usepackage[normalem]{ulem}
\usepackage{braket}

\usepackage{xcolor}
\usepackage{lipsum}

\newtheorem{theorem}{Theorem}

\newtheorem{example}{Example}

\newtheorem{proposition}{Proposition}
\newtheorem{remark}{Remark}

\DeclareMathOperator{\cA}{\mathcal{A}}

\DeclareMathOperator{\bR}{\mathbb{R}}

\DeclareMathOperator{\bP}{\mathbf{P}}
\DeclareMathOperator{\ind}{\mathbbm{1}}
\DeclareMathOperator{\bE}{\mathbf{E}}

\newcommand*\diff{\mathop{}\!\mathrm{d}}

\newcommand*\nnb{\nonumber}

\definecolor{sandy}{HTML}{E6E2AF}
\definecolor{stone}{HTML}{A7A37E}
\definecolor{beach}{HTML}{EFECCA}
\definecolor{ocean}{HTML}{046380}
\definecolor{diver}{HTML}{002F2F}

\definecolor{Firenze1}{HTML}{468966}
\definecolor{Firenze2}{HTML}{FFF0A5}
\definecolor{Firenze3}{HTML}{FFB03B}
\definecolor{Firenze4}{HTML}{B64926}
\definecolor{Firenze5}{HTML}{8E2800}

\definecolor{hongik}{HTML}{004498}

\definecolor{UT}{HTML}{CC5500}
\definecolor{airforceblue}{rgb}{0.36, 0.54, 0.66}
\definecolor{aqua}{rgb}{0.0, 1.0, 1.0}
\definecolor{awesome}{rgb}{1.0, 0.13, 0.32}
\definecolor{bostonuniversityred}{rgb}{0.8, 0.0, 0.0}
\definecolor{turquoise}{rgb}{0.19, 0.84, 0.78}
\definecolor{safetyorange}{rgb}{1.0, 0.4, 0.0}
\definecolor{Matlabblue}{rgb}{0, 0.4470, 0.7410}

\newcommand{\cc}[1]{\textcolor{black}{#1}}

\title{{LOS Coverage Area in Vehicular Networks with Cox-distributed Roadside Units and Relays}}

\author{Chang-Sik~Choi and François Baccelli
	\IEEEcompsocitemizethanks{\IEEEcompsocthanksitem 	{Chang-Sik Choi is with Hongik University, South Korea. (email: chang-sik.choi@hongik.ac.kr). François Baccelli is with Inria Paris and Telecom Paris, France. (email: francois.baccelli@inria.fr).}}
}

\begin{document}
	\maketitle 
	\begin{abstract}	
We develop an analytical framework to examine the line-of-sight (LOS) coverage area in vehicular networks with roadside units (RSU) and vehicle relays. In practical deployment scenarios, RSUs and vehicle relays are spatially correlated and we characterize this by employing Cox point processes to model the locations of RSUs and vehicle relays simultaneously. Leveraging the random blockage model, we model the LOS coverage area as Boolean models on these Cox point processes. The LOS coverage area is then evaluated by its area fraction. We show that relays can increase the area fraction of LOS coverage by nearly 50\% even when RSUs and relays are spatially correlated. By presenting a stochastic geometry model for a vehicular network with RSUs and relays and then by providing a tool to capture its LOS coverage, our work assesses the viability of vehicle relays for modern vehicular networks exploiting LOS coverage.
	\end{abstract}

	\section{Introduction}
	\subsection{Motivation and Related Work}
	Modern vehicular networks are envisioned to feature various applications such as cooperative driving, sensor data sharing, vehicle and pedestrian positioning, and Internet-of-Things (IoT) data sharing \cite{5510714,5888501,6702523,7992934,8082781,8255748,9184905}. In these applications, line-of-sight (LOS) signals play essential roles. For instance, in positioning applications, vehicles and pedestrians calculate their locations by exploiting the time differences of LOS signal arrivals \cite{fischer2014observed,8246850,8307478}. Similarly, a large amount of sensor data collected from infrastructure and advanced vehicles can be transmitted over LOS mmWave vehicle-to-all (V2X) communications \cite{7937682}. The fundamental limits of these applications are dictated by the network's LOS coverage.
		
\par To increase the LOS coverage of vehicular networks, modern vehicular networks have roadside units (RSUs) that provide the LOS signals to users on roads. However, due to the random obstacles or the complex motions of vehicles, the LOS signals from RSUs can be blocked, which undermines the sound operation of LOS-dependent applications. To ensure the LOS coverage to network users, many studies including \cite{7959158} have investigated various technologies. Among these, employing vehicle relays has been considered an effective and practical way to expand the LOS coverage of RSUs \cite{6240247,7891007,8417763,9685113,choi2022modeling}. Connected to the RSUs \cite{38836,38874}, these vehicle relays are capable of relaying the LOS signals from RSUs to users. For instance, in  \cite{9501036,9260163} the increment of the LOS coverage from vehicle relays was numerically evaluated; similarly, \cite{9119160} analyzed the reliability and latency of vehicular networks with vehicle relays;  an approach to choosing the best vehicle relay was explored in \cite{7949033}.

\par Leveraging stochastic geometry \cite{daley2007introduction,chiu2013stochastic,baccelli2010stochastic}, this paper analyzes the increment of the LOS coverage when RSUs employ vehicles as relays and these relays {expand} the LOS coverage of RSUs. In the context of stochastic geometry, the LOS coverage area or random blockage was studied in \cite{6290250,6840343,7848859,baccelli2001coverage,1392188}. Specifically,  in \cite{baccelli2001coverage,1392188}, random sets were considered on the Poisson point process to describe the LOS coverage of random transmitters. These Poisson-based approaches, however, can not capture the fact that RSUs and vehicle relays are on roads and so they are spatially correlated. Because of this spatial correlation, the LOS coverage jointly produced by them should overlap in space. To characterize the spatial correlation of RSUs and relays, we use Cox point processes \cite{morlot2012population,8340239,8357962,8419219} to simultaneously model the locations of RSUs and vehicles. Then, we model the LOS coverage as Boolean models on these RSU and vehicle relays. To the best of our knowledge, no existing work has analyzed such an overlap of LOS coverage, displayed by RSUs and vehicle relays.

	\subsection{Theoretical Contributions}
	\underline{Modeling for spatially correlated LOS coverage}:
This paper concerns the LOS coverage area that RSUs and their associated relays jointly produce on roads. We start by modeling the road layout as a Poisson line process. Then, we model RSUs and vehicles as Poisson point processes conditional on these roads. Eventually, the locations of RSUs and vehicles form Cox point processes. This modeling technique displays the geometric fact that both vehicles and RSUs are located on roads. Leveraging the random blockage model, we characterize the LOS coverage area as random rectangles centered on these Cox point processes. In contrast to the existing work on the blockage models where the spatial dependency between network components was overlooked \cite{6290250,6840343,7848859}, our work accounts for such a spatial dependency so that we can accurately evaluate the LOS coverage area obtained by vehicle relays. The LOS coverage area of this paper corresponds to the area where a network user can get LOS transmissions from RSUs or relays.

\underline{Derivation of the LOS coverage area}: 
Due to the spatial correlation of RSUs and relays, their LOS coverage overlaps. To account for this, we use the mean area fraction developed to examine the size of random sets\cite{chiu2013stochastic}. Leveraging the stationarity of LOS coverage present in the network and then getting the probability that the LOS coverage contains the origin, we derived formulas for the mean area fraction of the RSU LOS coverage and then that of the relays. Adapting key parameters from 3GPP, we evaluate the mean area fraction of the LOS coverage obtained under practical deployment scenarios. From these results, we conclude that relays increase LOS coverage by up to 50 percent. In other words, relays effectively expand the LOS coverage of RSUs and augment the service areas of the LOS-critical applications in modern vehicular networks, even though vehicle relays are spatially related to existing RSUs on roads. For completeness of the study, we perform large-scale Monte Carlo system-level simulations to experimentally obtain the mean area fraction of the LOS coverage for various cases. The simulation results confirm that the derived formulas are accurate. 

	\section{System Model}\label{S:2}
In this section, we present a spatial model for roads, RSUs, and vehicles. Then, we define the LOS coverage and then give the definition of the mean area fraction to analyze the size of the LOS coverage.

\subsection{Spatial Model}
	\begin{figure}
		\centering
		\includegraphics[width=1\linewidth]{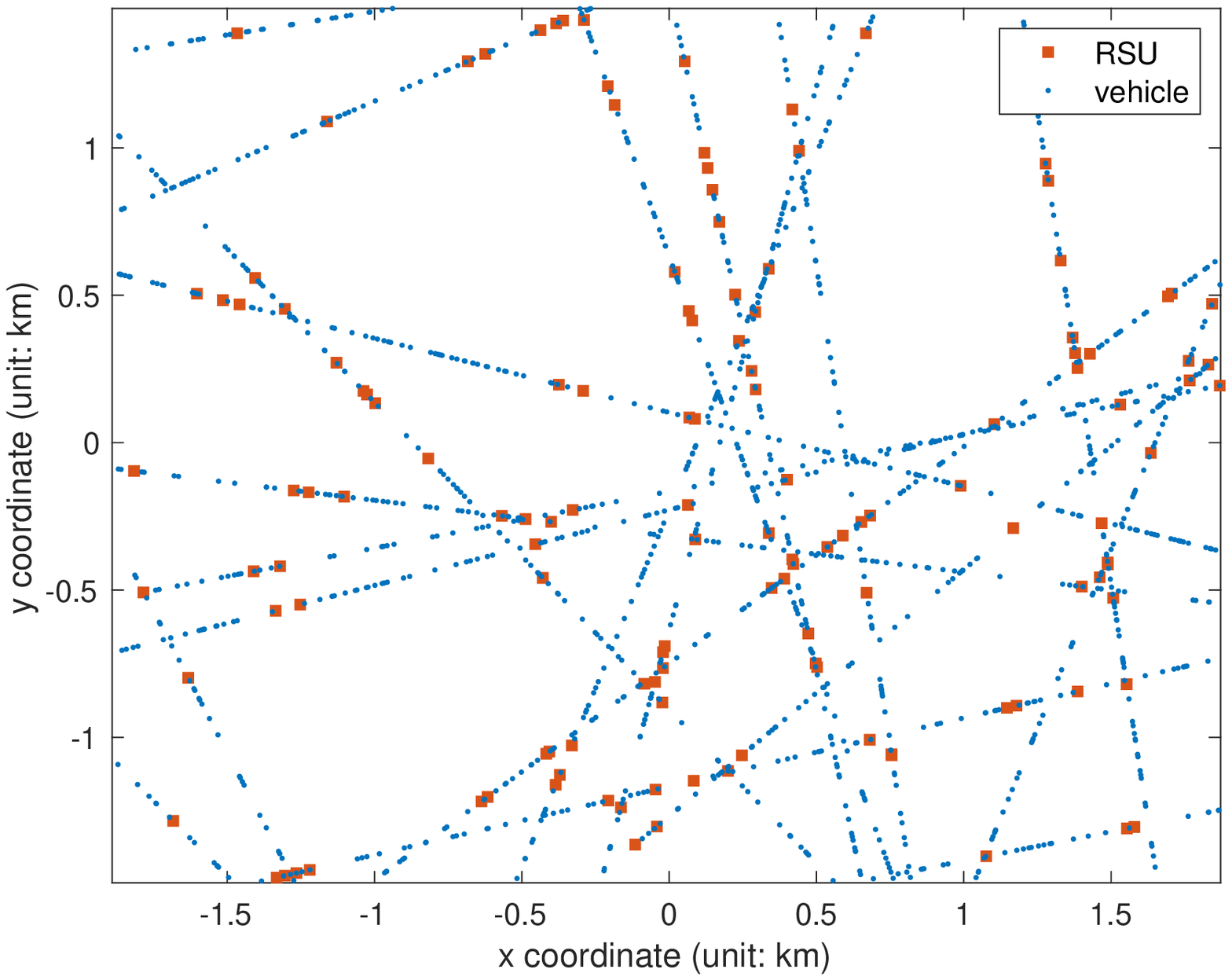}\\		
		\includegraphics[width=1\linewidth]{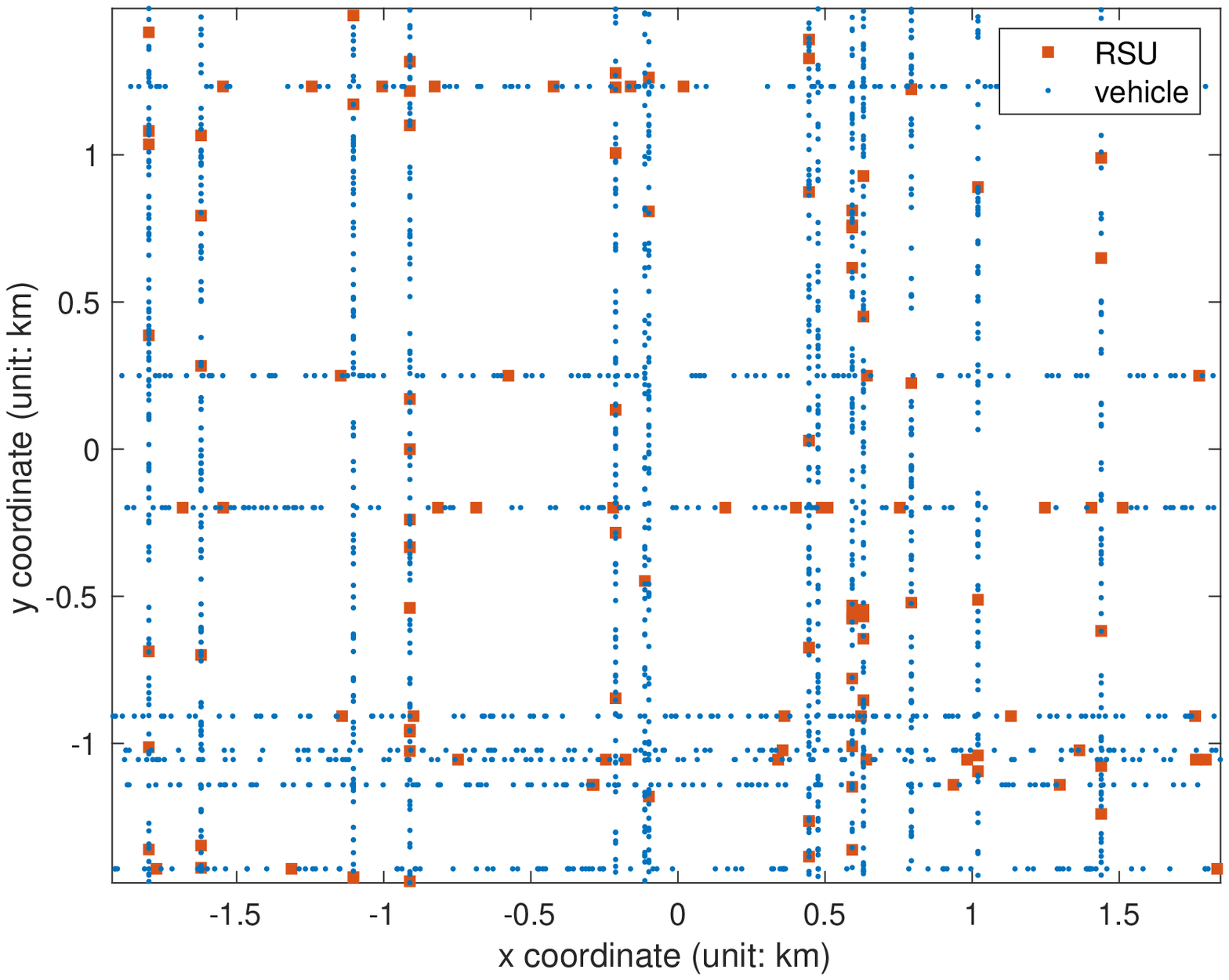}
		\caption{\cc{Simulation of the proposed model. Key parameters $ \lambda_l$, $ \mu $, and $ \mu_v $ are derived from practical deployment scenarios of 3GPP V2X evaluation methodology \cite{36885,37885}. We have $ \lambda_l=5 $/km, $ \mu=2 $/km, and $ \mu_v=25$/km.}}
		\label{fig:practicalcaserandom}
	\end{figure}

	\begin{figure}
		\centering
		\includegraphics[width=1\linewidth]{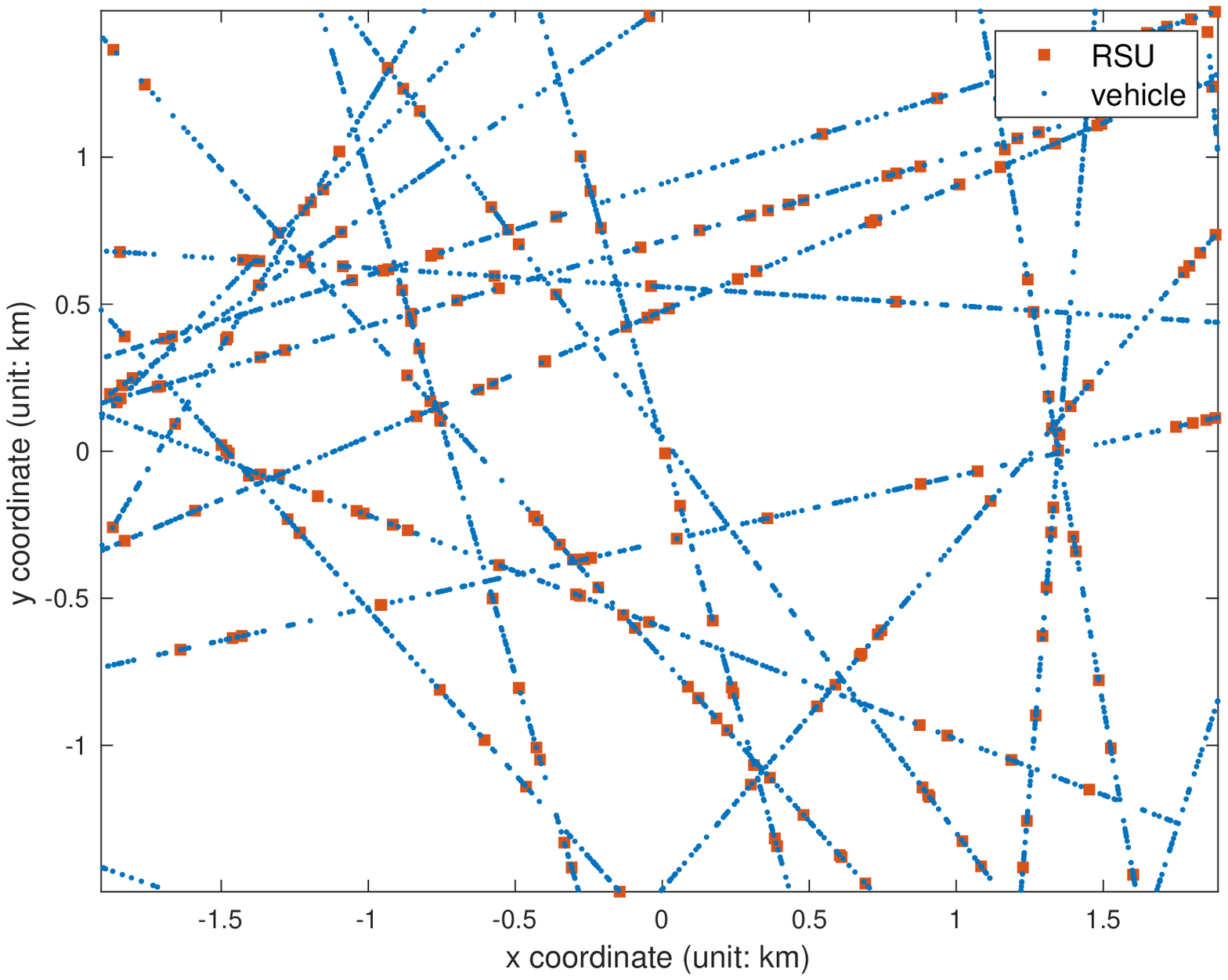}
		\includegraphics[width=1\linewidth]{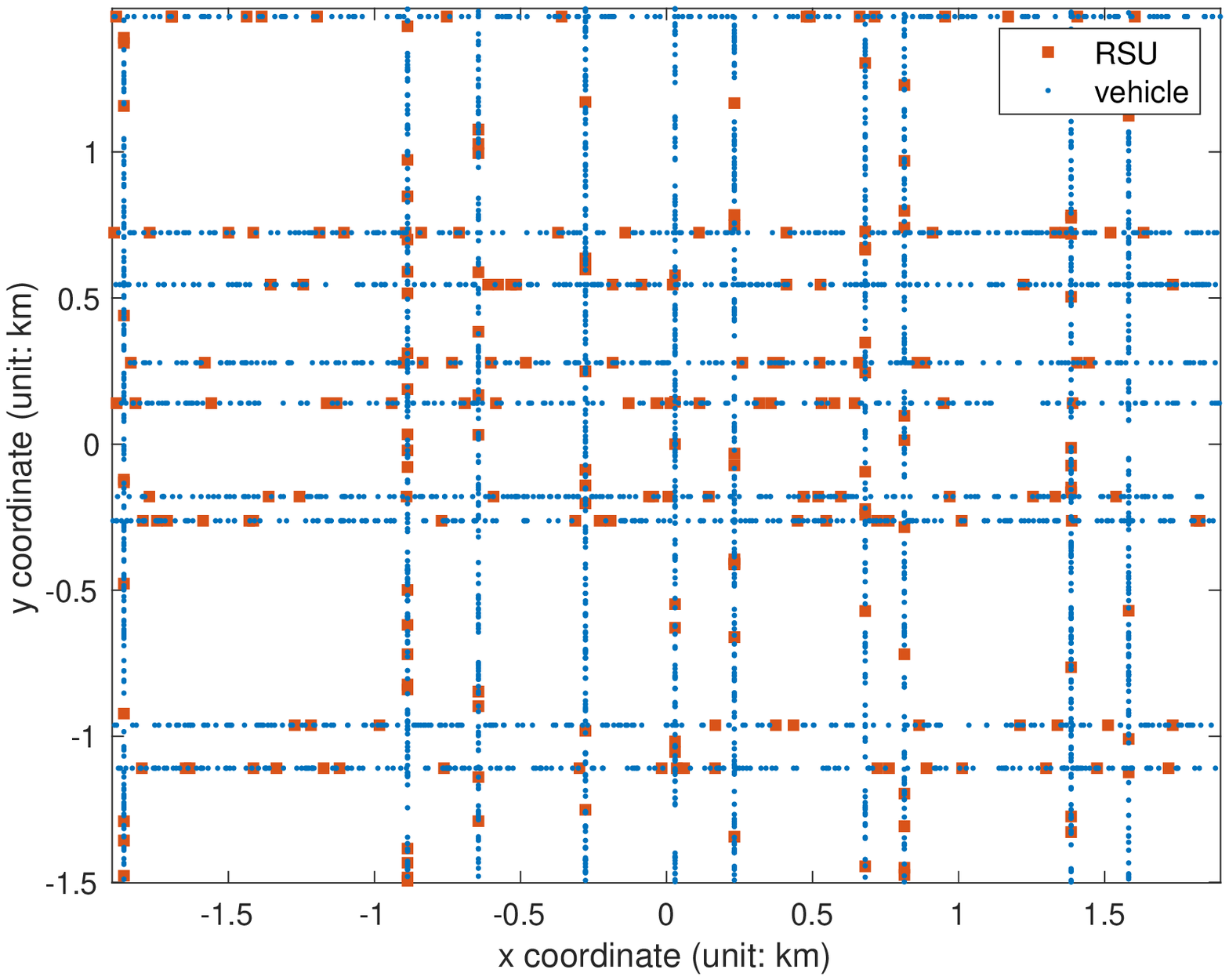}
		\caption{\cc{Simulation of the proposed model. Key parameters $ \lambda_l$, $ \mu $, and $ \mu_v $ are derived from practical deployment scenarios of 3GPP V2X evaluation methodology \cite{36885,37885}. We have $ \lambda_l=5 $/km, $ \mu=4 $/km, and $ \mu_v=50$/km.}}
		\label{fig:practicalm30msec}
	\end{figure}

	Using stochastic geometry \cite{daley2007introduction,chiu2013stochastic}, we model the set of roads as an isotropic Poisson line process, \cc{which has been widely used to model vehicular networks \cite{8340239,8357962,Choi2018Densification,8822637,9204398}}. The Poisson line process $ \Phi_l $ is generated by a Poisson point process $ \Xi $ of intensity $ \lambda_l/\pi $ on cylinder $ \mathbf{C} = \bR\times [0,\pi) $ where each point $ (r_i,\theta_i) $ of $ \Xi $ corresponds to an undirected line $ l(r_i,\theta_i) $ on $ \bR^2$.  \cc{Here, the first coordinate $ r_i $ is the distance from the origin to $ l(r_i,\theta_i) $ and the second coordinate $ \theta_i $ is the angle between $ l(r_i,\theta_i) $ and the $ x $-axis, measured in the counterclockwise direction. }

	\par Conditionally on $ \Phi_l$, the locations of RSUs are modeled as independent Poisson point processes of intensity $ \mu $. The RSU point process is given by $ 	\Phi = \sum_{(r_i,\theta_i) \in \Xi}\phi_{l(r_i,\theta_i)}. $	Similarly, conditional on the same line process $ \Phi_l, $ the locations of vehicles are modeled as independent Poisson point processes of intensity $ \mu_v $. Collectively, the vehicle point process is denoted by $ \Phi_v = \sum_{(r_i,\theta_i) \in \Xi}\psi_{l(r_i,\theta_i)} $. 
We assume that $ \mu_v \gg \mu $, namely the density of vehicles is much greater than the density of RSUs. 
\par 
	The locations of RSUs and the vehicles form stationary Cox point processes conditional on the same line process \cite{8419219,Choi2018Densification,choi2022modeling}. As a result, the proposed modeling technique accurately represents the fact that RSUs and vehicles are on roads and they are correlated in space. \par To address the motions of vehicles, this paper assumes that the vehicles move along their lines at a constant speed $ v $ \cite{36885}. Other mobility models are out of the scope of this paper.

\begin{figure}
	\centering
	\includegraphics[width=1\linewidth]{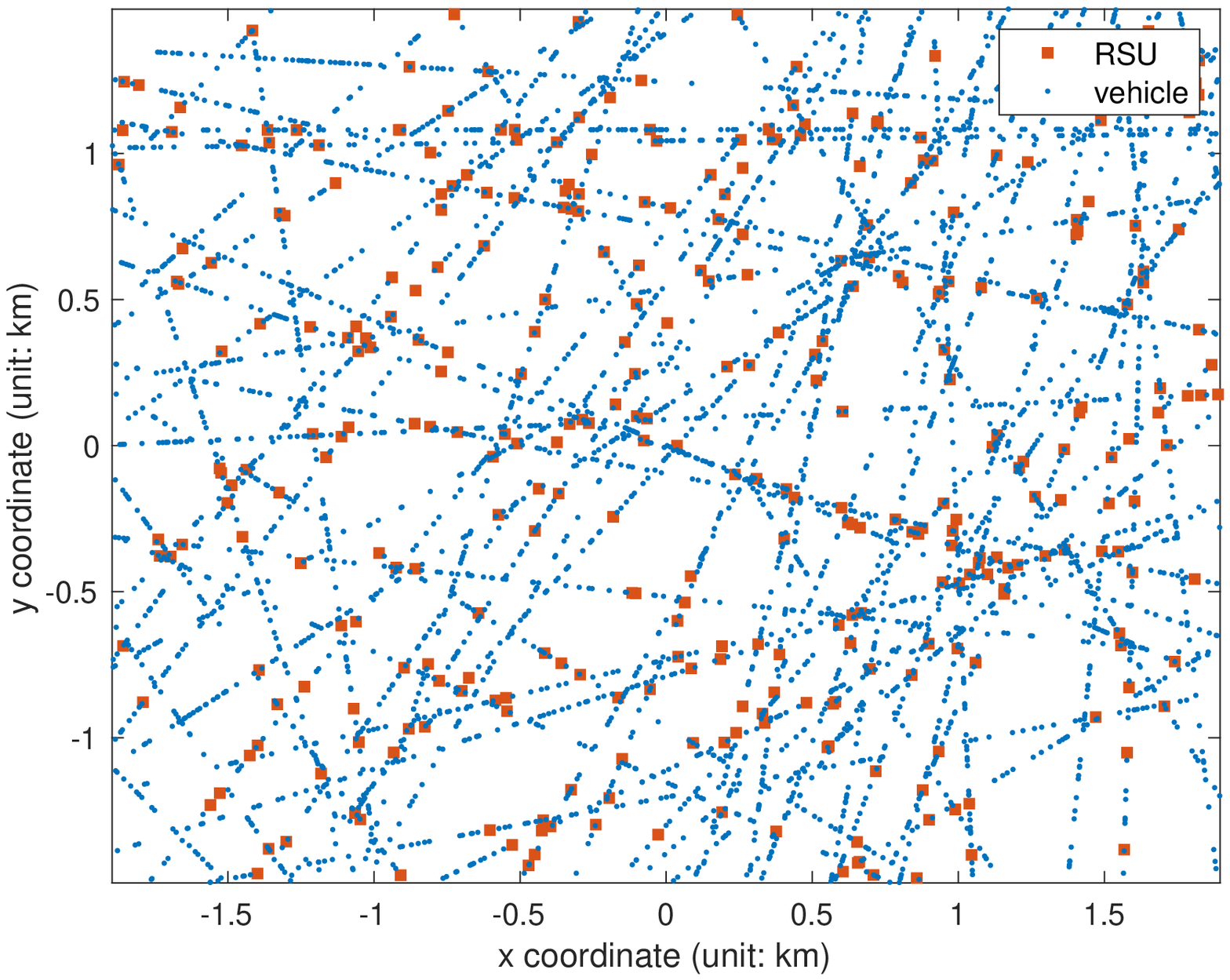}
		\includegraphics[width=1\linewidth]{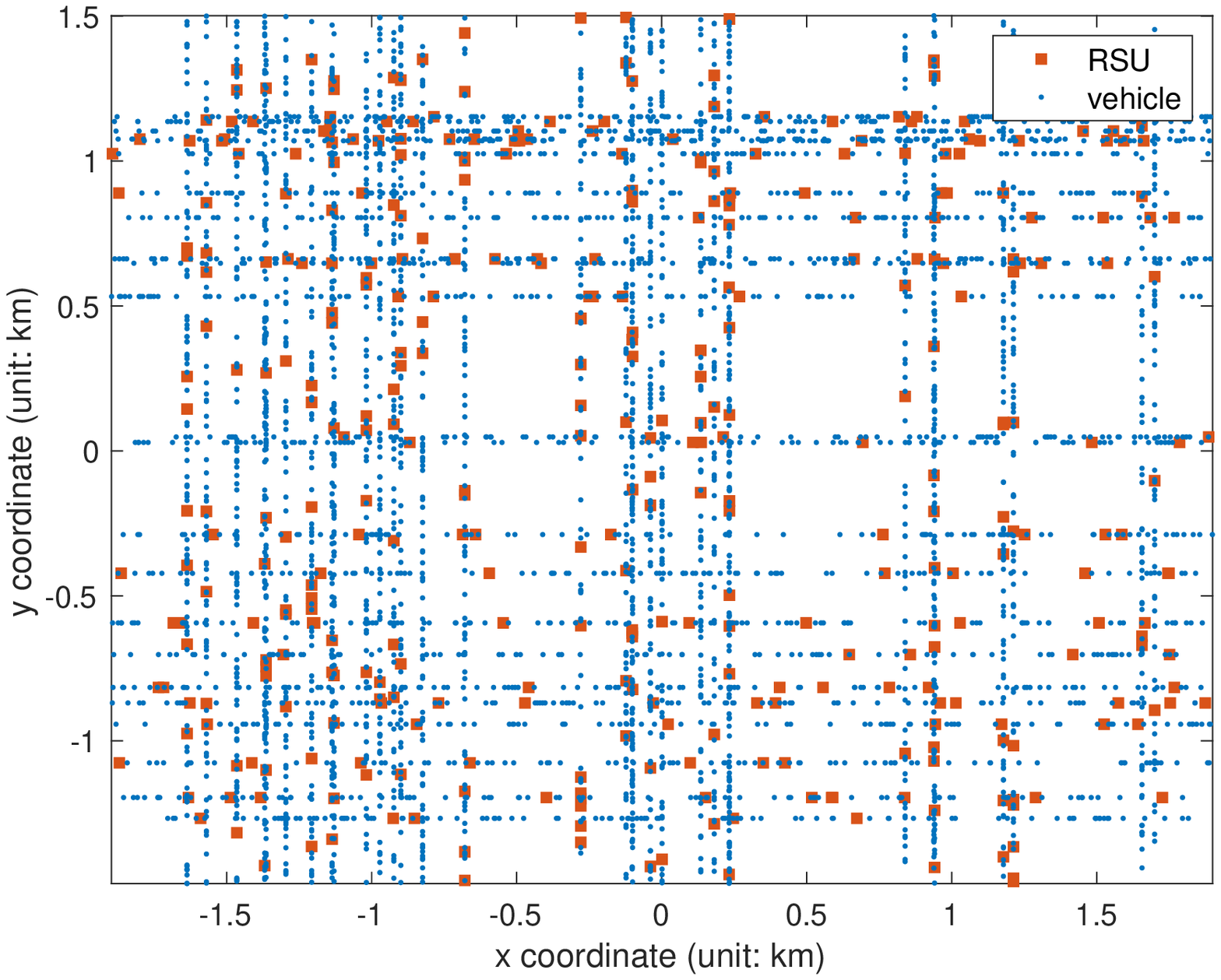}
	\caption{Simulation of the proposed model. We use $ \lambda_l =15$/km, $ \mu=2 $/km, $ v=20 $m/sec, and $ \mu_v =25$/km.}
	\label{fig:31}
\end{figure}

	\cc{Figs. \ref{fig:practicalcaserandom} -- \ref{fig:practicalm30msec} show that proposed spatial model representing practical deployment scenarios. Here, we use key parameters based on the 3GPP V2X evaluation methodology \cite{36885,37885}. Specifically, an urban scenario consists of road grids where each road grid is of size $ 430 $ meters $ \times  $ $ 250 $ meters and a simulation area of radius $ 1 $ km disk contains $ 10 $ road line segments \cite{36885}. Since the variable $ \lambda_{l} $ of this paper corresponds to the average number of roads in a disk of radius $ 0.5 $ km, we use $ \lambda_l  =5$/km in Figs. \ref{fig:practicalcaserandom} -- \ref{fig:practicalm30msec}. Similarly, since inter-RSU distance is given by $ 500 $ meters\cite{36885}, we use $ \mu = 2 $/km. The inter-vehicle distances are modeled as i.i.d. exponential random variables with mean $ 2v $ where $ v $ is the speed of vehicles (m/sec)\cite{36885}, we use $ v=10 $m/sec, and $ \mu_v = 50$/km. The Manhattan road layout of Fig. \ref{fig:practicalcaserandom} is produced by assuming that the angles of roads are either $0  $ or $ \pi/2 $ equally likely. Fig. \ref{fig:practicalm30msec} shows another example where inter-RSU distances is $ 250 $ meters\cite{36885}. In Fig. \ref{fig:practicalm30msec}, we use  $ \lambda_l=5 $/km, $ \mu=4 $/km, $v= 10 $ m/sec, and $ \mu_v = 50$/km. Note that our framework creates various road layouts and their vehicles by changing the value of $ \lambda_l $, $ \mu, $ and $ \mu_v $. To examine dense urban areas, we employ a higher value of $ \lambda_l$. Fig. \ref{fig:31} shows the proposed model with $ \lambda_l=15 $/km, $ \mu=4$/km, and $ \mu_v=25$/km.}
  
\begin{figure}
		\centering
		\includegraphics[width=.86\linewidth]{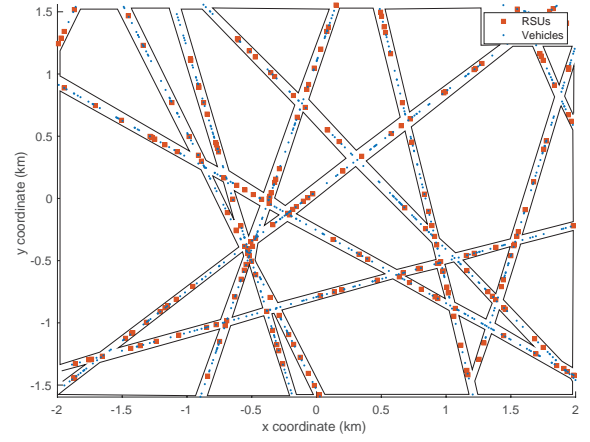}
		\caption{RSUs and vehicles are located on roads of width $ 75 $ meters. The polygons indicate LOS-inaccessible place such as building interiors.}
		\label{fig:rsusandvehiclesmodifiednofilled}
	\end{figure}

 \cc{To model the LOS coverage near road structure, we assume that each road is of width $  \eta  $ where RSUs or relays can provide the LOS signals to network users---such as pedestrians or smart sensors--- therein. Fig. \ref{fig:rsusandvehiclesmodifiednofilled} shows $ \eta = 75 $ meters.}

\begin{remark}
\cc{In practice, vehicles and RSUs are on the roads. We capture this by employing the vehicle and RSU Cox point processes jointly created under the same Poisson line process. Since the spatial distribution of transmitters dominates the LOS coverage, the proposed model leads to accurate estimates of the size of the LOS coverage. We will soon see that LOS coverage overlaps because RSUs and vehicle relays are spatially correlated. }
\end{remark}

	\subsection{Blockage, LOS Coverage, and Relay Selection}\label{S:2-B}
	To model the LOS coverage in vehicular networks, we leverage a random geometric model \cite{6840343,36885,37885,9759490}. Specifically, we assume that the LOS-blocking obstacles are located at distances $ W $ and $ V $ from any given transmitter and that $ W $ and $ V $ are i.i.d. exponential random variables with mean $ \gamma $. Since $ W $ and $ V $ are distances to LOS-blocking obstacles, they are also referred to as the LOS distances.
	
	\par
\begin{figure}
		\centering
		\includegraphics[width=1\linewidth]{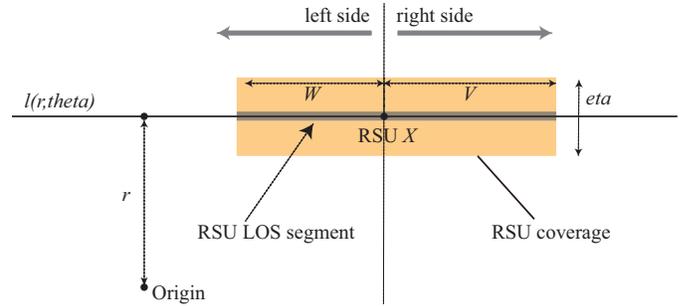}
		\caption{Illustration of the LOS segment of $ X $ and its LOS coverage. We denote by $ 0_i $ the location of the line $ l(r_i,\theta_i) $ closest to the origin. Other RSUs are omitted in this picture. }
		\label{fig:notation}
	\end{figure}

 Then, we define the LOS segment of an RSU as the line segment on which {vehicles of the same roads} are accessible to the LOS signal of the RSU. Specifically, the LOS segment of RSU $ X \in \phi_{l(r,\theta)} $ is defined by 
	\begin{equation}
		S_{X} = \begin{cases}
			\{x\in \text{left}(l(r,\theta);X) \ s.t. \|x - X\| < W_i \}\\
			\{x\in \text{right} (l(r,\theta);X) \ s.t. \|x - X\| < V_i \}
		\end{cases},
	\end{equation}
where $\text{left} (l(r,\theta);X) $ denotes the part of the line $ l(r,\theta) $ that is the left of the RSU $ X $ and  $\text{right} (l(r,\theta);X) $ denotes the part of the line $ l(r,\theta) $ that is the right of the RSU $ X $. Fig. \ref{fig:notation} shows the LOS segment of the RSU as one-dimensional segment on the line $ l(r,\theta) $.   
\par Then, we define the LOS coverage of RSU $ X $ as the \cc{place} where network users receive a LOS signal from the RSU $ X $. Leveraging the LOS segment of RSU $ X $ and finite width of roads in practice, the LOS coverage of RSU $ X $  is defined as a two-dimensional rectangle on the line $ l(r,\theta) $ whose width and length are $ \eta $ and the length of LOS segment of $ X $, respectively. The LOS coverage of RSU $ X $ is 
denoted by 
\begin{equation}
	C_{X}= S_{X} \times [- \eta /2,  \eta /2].
\end{equation}
This is the orthogonal product of two one-dimensional segments $ S_X $ and $ [-\eta/2,\eta/2] $. See Fig. \ref{fig:notation}.
\par
\begin{figure}
		\centering
		\includegraphics[width=0.9\linewidth]{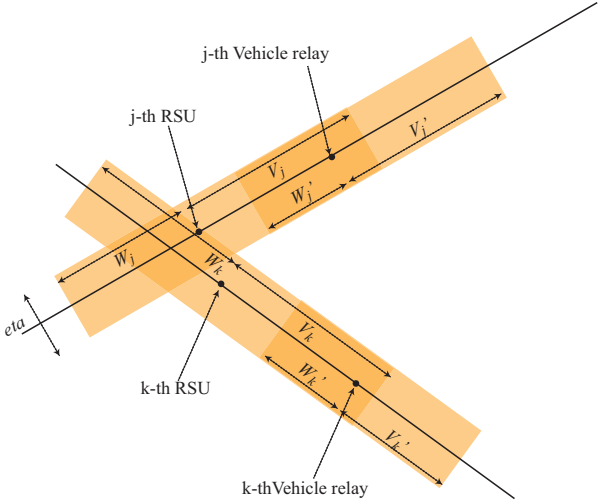}
		\caption{We use the subscripts to identify RSUs and relays. Here, the RSU coverage $ C_{X_j} $ and its relay coverage $ C_{Y_j} $ overlap due to the spatial dependency. Similarly, the RSU coverage of RSU $ X_j $ and RSU $ X_k $ overlaps. Arithmetic addition of LOS coverage areas overestimates the LOS coverage. See \ref{S:4-A} for a detailed explanation.}
		\label{fig:vehiclerelayv2rect}
	\end{figure}

 Similarly, the RSU LOS coverage is defined as the \cc{place where a network user gets at least one LOS signal from any RSU}. The RSU LOS coverage is given by 
	 \begin{equation}
		C_{\text{RSU}} =  \bigcup\limits_{X_i\in\Phi}C_{X_i} = \bigcup\limits_{X_i\in\Phi}\left(S_{X_i} \times [- \eta /2,  \eta /2] \right),\label{3}
	\end{equation} 	
where we use the {union} to quantify the LOS coverage of all RSUs since LOS coverage from distinct RSUs may overlap in space as shown in Fig. \ref{fig:vehiclerelayv2rect} . Since the LOS coverage is the collection of i.i.d. random sets centered on the RSU Cox point process, it is a Boolean (particle) model based on the RSU Cox point process. 

	In this paper, we focus on a scenario where each RSU selects one vehicle inside its LOS segment as a relay.  In other words, only vehicles in $ C_{\text{RSU}} $---the RSU LOS coverage---are considered as potential relay candidates. \cc{In practice, the proposed relay selection principle can be employed by V2X systems where vehicles in LOS with respect to (w.r.t.) at least one RSU are enabled to extend the LOS coverage of RSU.} For analytical tractability, we disregard the two following cases: (i) a vehicle is selected by more than two RSUs for their relays and (ii) there is no vehicle in the RSU LOS coverage. \cc{These two cases easily met in most practical use cases where the number of vehicles is greater than the number of RSUs \cite{36885}.} 
	
	As in the LOS coverage of RSUs, we assume that the selected relays have their own LOS coverage distances and that the relay LOS coverage is given by the union of the LOS coverage of all of these selected relays. Specifically, we define the locations of relays, their LOS segments, and their LOS coverage as follows:  
	\begin{align}
		\Phi' &=  \sum_{X_i\in\Phi}\text{Uniform}[\Phi_v(S_{X_i})]=\sum_{Y_i}\delta_{Y_i}\label{4},\\
		S_{Y_i} &=\begin{cases}
						\{x\in \text{left} (l(r,\theta);Y_i) | \|x - Y_i\| < W_i' \}\\
			\{x\in \text{right} (l(r,\theta);Y_i) | \|x - Y_i\| < V_i' \}
		\end{cases},\\
	C_{\text{Relay}} &=  \bigcup\limits_{Y_i\in\Phi'}C_{Y_i} = \bigcup\limits_{Y_i\in\Phi'}\left(S_{Y_i} \times [- \eta /2,  \eta /2] \right).
	\end{align}
Note that  the location of a selected relay is denoted by $ \text{Uniform}[\Phi_v(S_{X_i})]:$ a point selected uniformly random from the Poisson point process $ \Phi_v (S_{X_i}) $ where $ S_{X_i} $ is the LOS segment of the RSU $ X_i. $ As in the LOS distances defined for RSUs, we assume that $ W_i' $ and $ V_i' $ follow i.i.d. exponential random variables with mean $ \gamma $.

Besides, under the assumption that $ \mu_v\gg \mu, $ we have  $ \text{Uniform}[\Phi_v(S_{X_i})] \approxeq \text{Uniform}(S_{X_i})$, a random point uniformly distributed on the segment $ S_{X_i}.$ Consequently, we have 
	\begin{equation}
			\Phi' =\sum_{Y_i}\delta_{Y_i}=  \sum_{X_i\in\Phi}\text{Uniform}(S_{X_i}).\label{7}
	\end{equation}

Finally, we define the RSU-plus-relay LOS coverage as the area where the network users are in LOS w.r.t. at least one LOS signal from either an RSU transmitter or a relay transmitter. Using Eqs. \eqref{4} and \eqref{7}, the relay LOS coverage $ C_{\text{Relay}} $ and the RSU-plus-relay LOS coverage $ C_{\text{RSU+relay}} $   are defined as follows:
	 \begin{align}
	 	C_{\text{Relay}} &= \bigcup\limits_{X_i\in\Phi}\left( S_{\text{Unif}(S_{X_i})} \times [- \eta /2,  \eta /2] \right),\\
	C_{\text{RSU+relay}} &=  C_{\text{RSU}} \cup C_{\text{Relay}}  \nnb\\
	&= \bigcup\limits_{X_i\in\Phi}\left((S_{X_i}\cup S_{\text{Unif}(S_{X_i})}) \times [- \eta /2,  \eta /2] \right),
\end{align} 	
Because of the spatial correlation between RSUs and relays, there should be overlap of LOS coverage of RSUs and relays. Fig \ref{fig:vehiclerelayv2rect}  shows such an overlap of LOS coverage in vehicular networks.

\begin{example}
	This paper assimilates the network users' perspectives to define LOS coverage. Specifically, the LOS coverage is the place where a network user gets at least one LOS signal from any transmitter. In modern vehicular applications such as positioning or mmWave communications, the existence of a LOS signal is necessary for the operation of those applications; thus, the LOS coverage indicates the area where network operators can implement such applications.
\end{example}

	\subsection{Performance Metric}
	To derive the area of the LOS coverage, we use the mean area fraction \cite{daley2007introduction,baccelli2010stochastic,chiu2013stochastic}. The mean area fraction of a Boolean model $ \cA $ \cite{daley2007introduction,schneider2008stochastic} is the ratio of the mean area of $ \cA $ to the mean area of $ \bR^2 $. The mean area fraction of $ \cA $ takes a value between zero and one: it is one if $ \cA $ covers the entire plane; it is zero if the area of $ \cA $ is negligible compared to that of $ \bR^2 $. 
	
	The mean area fraction of the RSU LOS coverage is 
	\begin{align}
		 \nu(C_{\text{RSU}}) \!= \!\bE\left[\!\int_{C_0}\ind_{C_{\text{RSU}}\cap C_0}(x) \diff x \right]\!=\!\lim_{r\to\infty}\!\frac{\int_{B_0(r)\cap C_{\text{RSU}}} \ind_{x} \diff x }{\int_{B_0(r)} \ind_{x} \diff x }\nnb,
	\end{align}
 where $ C_0  $ is a unit square $ \bR^2 $ and $ B_0(r) $ is the ball of radius $ r $ centered at the origin. 
 \par The mean area fraction of RSU-plus-relay LOS coverage is 
		\begin{align}
	\nu(C_{\text{RSU+relay}})&=\nu(C_{\text{RSU}}\cup C_{\text{Relay}}) \nnb\\
	&= \bE\left[\int_{C_0}\ind_{( C_{\text{RSU}}\cup  C_{\text{Relay}})\cap C_0}(x) \diff x \right]\nnb\\
	&=\lim_{r\to\infty}\frac{\int_{B_0(r)\cap (C_{\text{RSU}}\cup C_{\text{Relay}})} \ind_{x} \diff x }{\int_{B_0(r)} \ind_{x} \diff x }.
\end{align}	
\cc{In the proposed model, the LOS coverage corresponds to the area where LOS-dependent applications such as mmWave communications or vehicle positioning are operational. As a result, the increment of the mean area fraction of the LOS coverage area directly relates to the increment of the service area of these applications thanks to the newly added relays. In this paper, we will assess the benefits of employing vehicle relays by getting the difference between $ \nu(C_{\text{RSU+relay}}) $ and $ \nu(C_{\text{RSU}}). $  }


 	\section{Main Results}\label{S:3}
 	In this section, we analyze the mean area fraction of the LOS coverage. 
 	\subsection{Stationarity of the LOS Coverage}
 	\begin{proposition}\label{L:1}
 		The RSU coverage $  C_{\text{RSU}} $, the vehicle relay coverage $  C_{\text{Relay}}, $ and the RSU-plus-relay coverage $  C_{\text{RSU+relay}} $ are stationary.  
 	\end{proposition}
 	\begin{IEEEproof}
 		The RSU coverage $  C_{\text{RSU}}  $ is the union of i.i.d. random sets centered on the RSU Cox point process. Since the RSU point process is also stationary (translation invariant) \cite{8419219}, the RSU LOS coverage $  C_{\text{RSU}}  $ is stationary \cite{baccelli2010stochastic}. 
 		
 		\par Moreover, since the relay point process $ \Phi' $ is stationary and its LOS coverage $  C_{\text{Relay}} $ is the union of i.i.d. random sets, the relay LOS coverage $ C_{\text{Relay}} $ is stationary. In a similar way, we have that the RSU-plus-relay LOS coverage $  C_{\text{RSU+relay}} $ is stationary.  
 	\end{IEEEproof}
 		\subsection{RSU LOS Coverage}
 	\begin{theorem}\label{T:1}
 		The mean area fraction of RSU coverage is 
 		\begin{equation}
\nu( C_{\text{RSU}} )=1-\exp\left(-\lambda_l  \eta  \left(1- e^{-2\mu \gamma}\right)\right).\label{eq:T1}
 		\end{equation}
 	\end{theorem}
 
  	 \begin{figure*}
 	\begin{equation}\label{eq:Theorem1}
 		1-\exp\left(-\lambda_l{ \eta }\left(1-\exp\left(-\mu\int_{-\infty}^{\infty}\left\{e^{-\frac{|x|}{\gamma}}-\int_{0}^{\infty}\int_{0}^{\infty}\int_{-w}^{v}\frac{1-e^{-\frac{|x+y|}{\gamma}}}{w+v} \frac{e^{-\frac{w+v}{\gamma}}}{\gamma^2}\diff y\diff w \diff v\right\}\diff x\right) \right)\right).
 	\end{equation}
 	\vspace{-2mm}
 	\\\rule{\textwidth}{0.3pt}
 \end{figure*}

 	\begin{IEEEproof}
 		From the stationarity of RSU coverage, the mean area fraction is equal to the probability that the origin is contained in the RSU coverage region \cite{chiu2013stochastic}.  Therefore, we have  $\nu( C_{\text{RSU}} ) \equiv \bP(0\in C_{\text{RSU}} )=1-\bP(0\notin  C_{\text{RSU}} ). $

		Let $ 0_i $ be the point on the line $ l(r_i,\theta_i) $ closest to the origin. Then, for a given line $ l(r_i,\theta_i) $ the typical point at the origin is not contained in the LOS coverage of the RSUs on the line if (i) $ W_j < |X_j - 0_i |$ for all the RSUs $ X_j $ on the right side of $ 0_i $ and (ii) $ V_j<|X_j-0_i| $ for all the RSUs $ X_j $ on the left side of $ 0_i $.  
		
		In addition, for any line $ l(r_i,\theta_i) $, the LOS coverage of RSUs on those lines does not contain the origin if and only if the distances from the origin to the line is greater than $ \eta/2. $  Then, we use the facts that that the lines of the Poisson line process are independent, that $ W_j$ and $ V_j $ are i.i.d., and that the locations of RSUs on each line is a Poisson point process of intensity $ \mu $. The mean area fraction of the RSU LOS coverage is given by 	
	\begin{align}
 		&\bE_{\Xi}\left[\prod_{r_i,\theta_i\in\Xi}^{|r_i|< \eta /2}\bE\left[\prod_{{X}_j\in\phi_{l(r_i,\theta_i)}}\bE\left[\left.\ind_{W<|X_j-0_i| }\right| \Phi_l,\phi\right]\right]\right]\nnb\\
 		&=\bE_{\Xi}\left[\prod_{r_i,\theta_i\in\Xi}^{|r_i|< \eta /2}\bE\left[\prod_{{T}_j\in\phi}\bE\left[\left.\ind_{W<|{T}_j| }\right| \Phi_l,\phi\right]\right]\right],
	\end{align}
where we use the Poisson property of $ \phi_{l(r_i,\theta_i)} $ to obtain 
\[ \sum_{X_j\in\phi_{l(r_i,\theta_i)}}\delta_{X_j-0_i} \equiv \sum_{T_j\in\phi}\delta_{T_j},\] 
where $ \phi $ is a Poisson point process of intensity $ \mu $ on $ \bR $. Therefore, we have 
  	\begin{align}
 		\bP(0\notin C_{\text{RSU}} )
 		&=\bE_{\Xi}\left[\prod_{r_i,\theta_i\in\Xi}^{|r_i|< \eta /2}\bE\left[\prod_{{T}_j\in\phi}\bE\left[\left.\ind_{W<|{T}_j| }\right| \Phi_l,\phi\right]\right]\right]\nnb\\
 		&=\bE_{\Xi}\left[\prod_{r_i,\theta_i\in\Xi}^{|r_i|< \eta /2} \bE\left[\left.\prod_{{T}_j\in\phi} 1- e^{-\frac{|T_j|}{\gamma}}\right|\Phi_l\right]\right]\nnb\\ 		
 		&=\bE_{\Xi}\left[\prod_{r_i,\theta_i\in\Xi}^{|r_i|< \eta /2}\exp\left(-2\mu \int_{0}^{\infty}e^{{-\frac{t}{\gamma}}}\diff t\right)\right]\nnb\\
 		&=\bE_{\Xi}\left[\prod_{r_i,\theta_i\in\Xi}^{|r_i|< \eta /2}\exp\left(-2\mu \gamma \right)\right]\nnb\\
 		 &=\exp\left(-\lambda_l  \eta  \left(1- e^{-2\mu \gamma}\right)\right),
\end{align}
where we use the total independence property of the Poisson point processes $ \phi $ and $ \Xi $, and their probability generating functionals \cite{baccelli2010stochastic}.  
 	\end{IEEEproof}
 
 \begin{figure}
 	\centering
 	\includegraphics[width=0.9\linewidth]{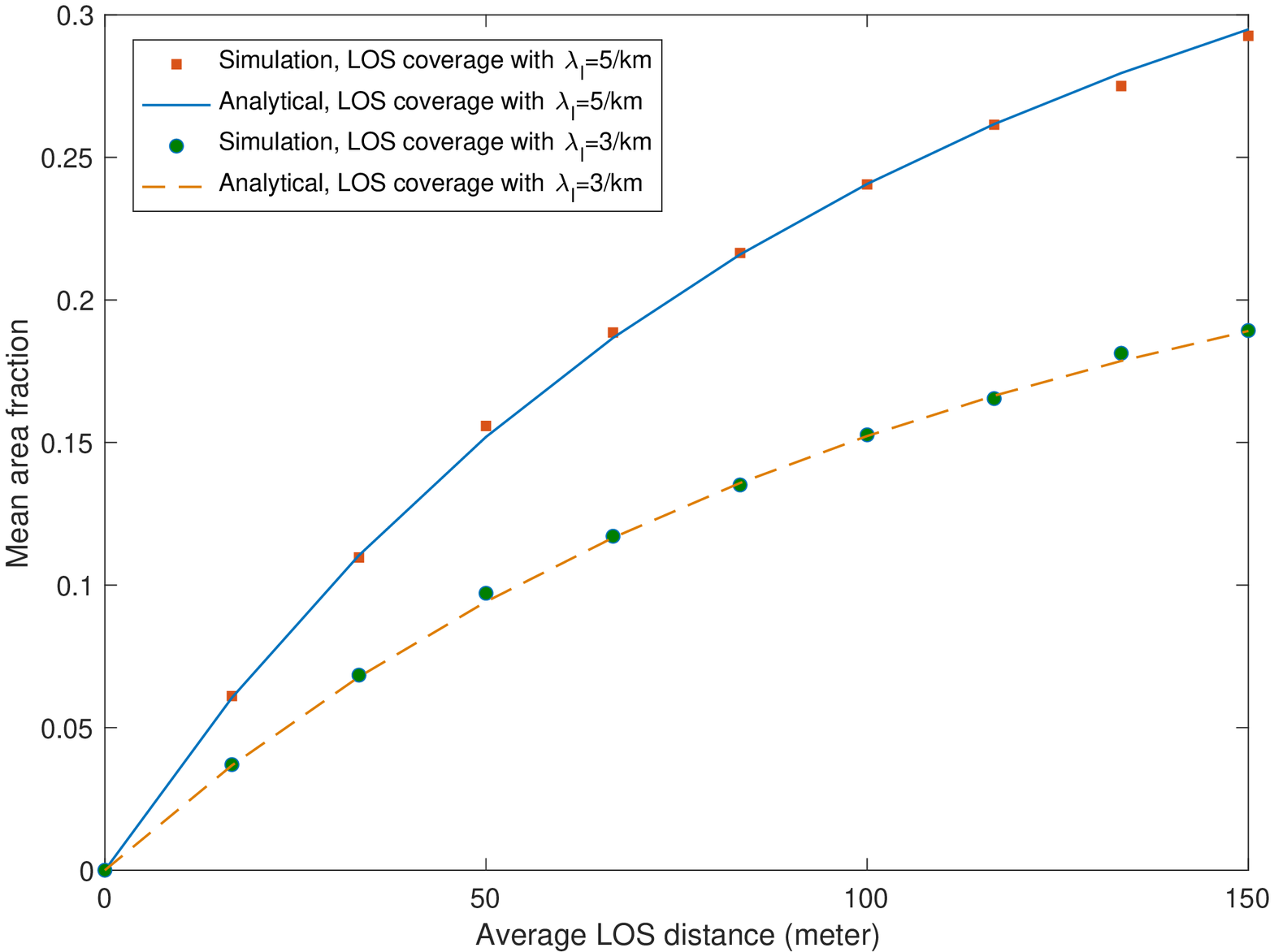}
 	\caption{{The mean area fraction of the RSU LOS coverage with $ \mu=4 $/km and $ \eta=100 $ meters. }}
 	\label{fig:eta100lambdal53mu4}
 \end{figure}
 
 \begin{figure}
 	\centering
 	\includegraphics[width=0.9\linewidth]{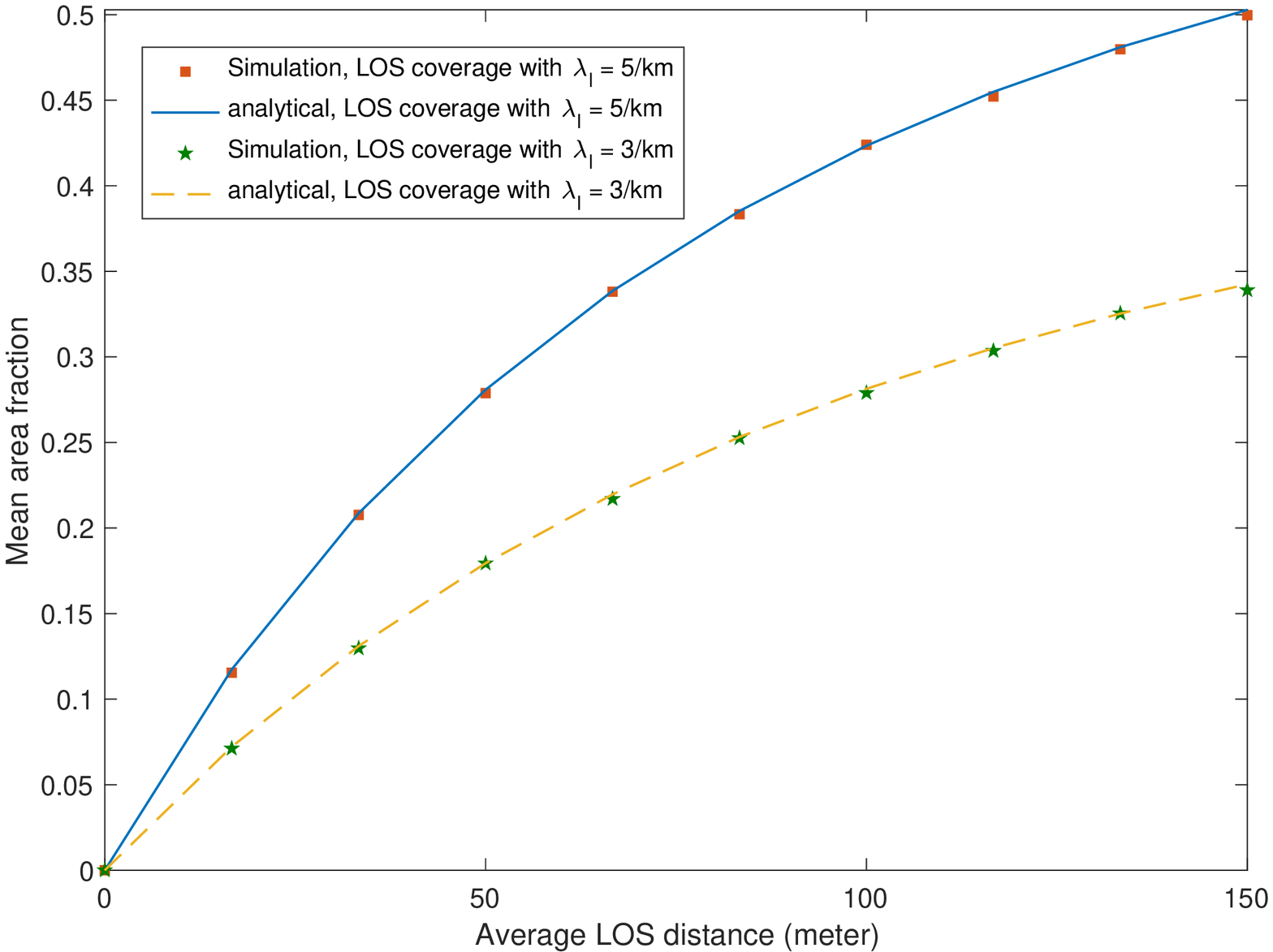}
 	\caption{The mean area fraction of the RSU LOS coverage with $ \mu=4 $/km and $ \eta=200 $ meters.}
 	\label{fig:rsucoveragesimvsanaly}
 \end{figure}
Figs. \ref{fig:eta100lambdal53mu4} and \ref{fig:rsucoveragesimvsanaly} show the mean area fraction of the RSU coverage as $ \gamma $---the average LOS distance---increases. They display that the derived formula accurately matches the simulation results for both $ \eta=100 $ and $ \eta=200 $. We learn that as $ \gamma  $ increases, $ \nu(C_{\text{RSU}}) $ also increases.  For instance, when $ \lambda_l=3/\text{km} $ and $ \gamma=50 $ meters, we have $ \nu(C_{\text{RSU}})  \approxeq  0.1 $; on the other hand, when $ \gamma=150 $ meters, we have $ \nu(C_{\text{RSU}})\approxeq 0.15.  $ Roughly, as $ \gamma $ triples, $ \nu(C_{\text{RSU}}) $ increases only by $ 50 $\%. This occurs because the RSUs are spatially correlated with each other and when they are close to each other their LOS coverage overlaps.  Note the increment of $ \gamma $ has a diminishing return on the increment of the area fraction because the size of overlap also increases. It is important to mention that the maximum of the mean area fraction of the LOS coverage is the area fraction of the all roads in the network. It is $ \nu(\Phi_l\oplus B_0(\eta/2))=1-e^{-\lambda_l \eta } $  \cite{8419219} where $ \lambda_l $ is the road density and $ \eta $ is the width of roads. For network instances studied in Figs. \ref{fig:eta100lambdal53mu4} and \ref{fig:rsucoveragesimvsanaly}, the corresponding mean area fractions of all roads are
 \begin{align}
 	\nu(\Phi_l\oplus B_0(\eta/2))= 		\begin{cases} 
 		 			0.39 & \lambda_l=5/\text{km}, \eta=100 \text{ m},\\
 		0.26 & \lambda_l=3/\text{km},  \eta=100 \text{ m},\\
 		0.63,  & \lambda_l=5/\text{km} , \eta=200 \text{ m}, \\
 		0.45, &   \lambda_l=3/\text{km},  \eta=200 \text{ m},
 	\end{cases}
 \end{align}
respectively.
\begin{remark}
To get the simulation results, we measure the mean area fraction of the LOS coverage set by the Monte Carlo method. Specifically, for each simulation instance, a Poisson line process is simulated on a disk of a radius $ 10  $km. Then, the RSUs are simulated on these lines. Based on $ \gamma $, the LOS coverage of all these  RSUs is created as rectangles centered on these RSUs. Then, by counting whether the origin is contained by the simulated LOS coverage or not for each instance, we obtain the probability that the origin is contained by the LOS coverage. For smoother results, we have more than $ 10^5 $ simulation instances.
\end{remark}

\begin{figure}
	\centering
	\includegraphics[width=1\linewidth]{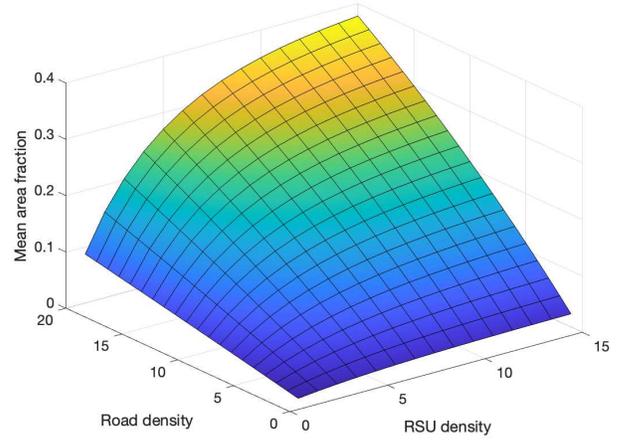}
	\caption{The mean area fractions of the LOS coverage with $ \gamma = 100 $ meters and $ \eta =25 $ meters.}
	\label{fig:eta25meter}
\end{figure}
\begin{figure}
	\centering
	\includegraphics[width=1\linewidth]{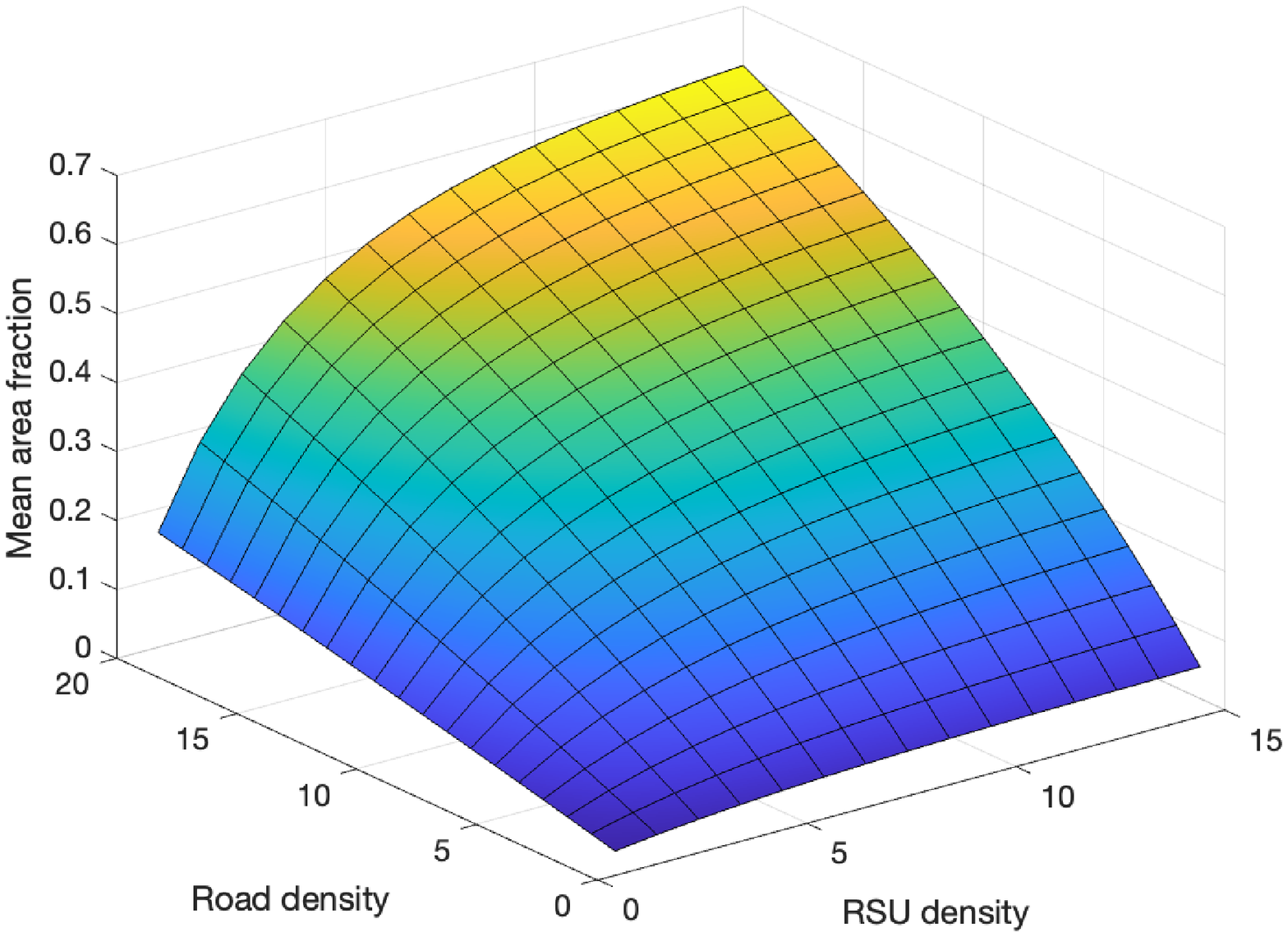}
	\caption{The mean area fractions of the LOS coverage with $ \gamma = 100 $ meters and $ \eta =50 $ meters.}
	\label{fig:eta50meters}
\end{figure}

\begin{figure}
	\centering
	\includegraphics[width=1\linewidth]{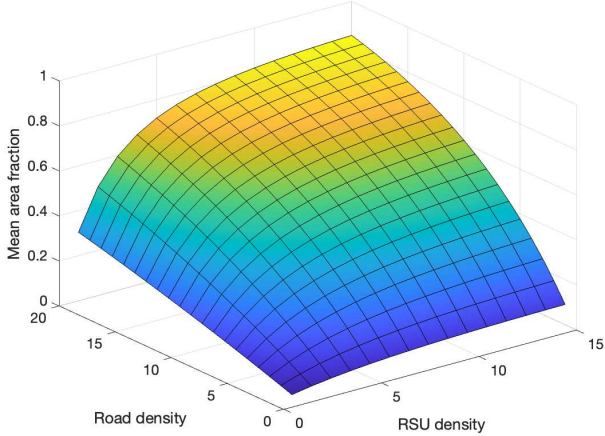}
	\caption{The mean area fractions of the LOS coverage with $ \gamma = 100 $ meters and $ \eta =100 $ meters.}
	\label{fig:gap2d1deta100exact}
\end{figure}

Figs. \ref{fig:eta25meter} -- \ref{fig:gap2d1deta100exact} illustrate the mean area fraction of the RSU LOS coverage when the road density $ \lambda_l $ and the RSU density $ \mu  $ vary simultaneously. The units of road density and RSU density are meters.

\subsection{RSU-plus-relay LOS Coverage}	
This section presents the main result: the mean area fraction of the LOS coverage created by both RSUs and their relays.

 	\begin{theorem}\label{T:2}
 		The mean area fraction of the RSU-plus-relay LOS coverage is given by Eq. \eqref{eq:Theorem1}.
 	\end{theorem}
 	\begin{IEEEproof}
 		\cc{The mean area fraction of the RSU-plus-relay LOS coverage is given by the probability that the origin is contained by the RSU-plus-relay LOS coverage \cite{chiu2013stochastic}. Let $ \ind_{A} $ denote an indicator function that takes one if the event $ A $ occurs and zero if an event $ A $ does not occur. The mean area fraction of the RSU-plus-relay LOS coverage is given by		$ 	\nu(C_{\text{RSU+relay}})=\bE[\ind_{0\in C_{\text{RSU+relay}}}]. $
 	Conditional on the line process $ \Phi_l $, we have }
 		\begin{align}
&\bE[\ind_{0\notin C_{\text{RSU+relay}}}]\nnb\\
 				&=\!\bE\left[\bE\left[\left.\prod_{\phi_i\in\Phi_l}\ind_{0\notin C_{\text{RSU}} \cup C_{\text{Relay}}}\right|\Phi_l\right]\right]\nnb\\
 			&=\bE\left[\prod_{r_i\in\Phi_l}\bE\left[\ind_{0\notin C_{\text{RSU}} \cup C_{\text{Relay}}}|\Phi_l\right]\right],\nnb\\
 			&=\bE\left[\prod_{r_i\in\Phi_l}\bE\left[\prod_{t_j\in\phi}\bE\left[\ind_{0\notin C_{\text{RSU}} \cup C_{\text{Relay}}}|\Phi_l,\phi\right]\right]\right],
 		\end{align}
 	where we use the same technique as in the proof of Theorem \ref{T:1}. Furthermore, conditional on the LOS distances, namely $ W_j $ and $ V_j $, the mean area fraction of the RSU-plus-relay LOS coverage is given by 
 	\begin{align}
 		\bE\left[\prod_{r_i\in\Phi_l}\bE\left[\prod_{t_j\in\phi}\bE\left[\bE\left[\ind_{0\notin C_{\text{RSU}} \cup C_{\text{Relay}}}|\Phi_l,\phi,W,V\right]\right]\right]\right],\nnb
 	\end{align} 
 where we drop the subscripts of the variables $ W $ and $ V $ for a simpler expression.
Then,	leveraging a simple set theory and the property of the indicator function, we have 
 	\begin{align}
 		\ind_{0\notin C_{\text{RSU}}  \cup  C_{\text{Relay}}} = \ind_{0\notin C_{\text{RSU}} } + \ind_{0\notin C_{\text{Relay}} \cap  C_{\text{RSU}} ^c}.
 	\end{align}
 The first term is a measurable function of the random variables $ \Phi_l,\phi,W, $ and $V $.  Therefore, the mean area fraction is 
 	\begin{align}
	&\bE\left[\prod_{\phi_i\in\Phi_l}\bE\left[\prod_{t_j\in\phi}\bE\left[\ind_{0\notin C_{\text{RSU}} }\right.\right.\right.\nnb\\
	&\hspace{2.5cm}\left.\left.\left.+\underbrace{\bE\left[\ind_{0\notin C_{\text{Relay}}\cap  C_{\text{RSU}} ^c}|\Phi_l,\phi,W,V\right]}_{(\text{a})}\right] \right]\right].\nnb
 	\end{align}
The origin is not contained in $  C_{\text{Relay}} \cap  C_{\text{RSU}} ^c $ if neither the LOS segments centered on the left-hand side of $ 0_i $ nor the LOS segments on the right-hand side of $ 0_i $ contains the origin, for all lines in the network. Let $ Y_j $ denote the location of the selected relay associated with the $ j $-th RSU. As in the proof of Theorem \ref{T:1}, the term (a) is given by 
\begin{align}
	(\text{a}) &=\begin{cases}
		\bE\left[\left.\ind_{W < |Y_j| }\right| \Phi_l,\phi,W,V\right], &Y_j\in\text{right}(l(r_i,\theta_i);0_i)\nnb\\ 
		\bE\left[\left.\ind_{V < |Y_j| }\right| \Phi_l,\phi,W,V\right],&Y_j\in\text{left}(l(r_i,\theta_i);0_i)\nnb\\ 
	\end{cases}\\
	&=\bE_{Y_j}\left[\left.  1-e^{-\frac{|Y_j|}{\gamma}} \right|\Phi_l,\phi,W,V\right]\nnb\\
	&=\int_{-w}^{v}\frac{1-\exp\left({-\frac{|t_j +y| }{\gamma}}\right)}{w+v}\diff y ,
\end{align}
where we use the facts that (i) the probability density function of the exponential random variable is $ f(x)= 1-\exp(-x/\gamma) $ and (ii)  the location of the relay $ Y_j $ is uniformly distributed within $ [t_j-w,t_j+v] $, where $ t_j $ is the location of the $ j $-th RSU and $ w $ and $ v $ are the LOS distance from the RSU. Here $ w $ and $ v $ are the sample values of the i.i.d. exponential random variables $ W $ and $ V $, respectively.

As a result, by deconditioning w.r.t. $ W $ and $ V $, the mean area fraction of RSU-plus-relay LOS coverage is given by 
\begin{align}
	&\bE\left[\prod_{r_i\in\Phi_l}\bE\left[\prod_{t_j\in\phi} \left(1-e^{-\frac{|t_j|}{\gamma}}\right.\right.\right.\nnb\\
	&\hspace{5mm}\left.\left.\left.+\int_{0}^{\infty}\!\!\int_{0}^{\infty}\!\!\left(\int_{-w}^{v}\frac{1-e^{-\frac{|t_j+y|}{\gamma}}}{w+v}\right) \frac{e^{-\frac{w+v}{\gamma}}}{\gamma^2} \diff y \diff w \diff v  \right)\right]\right]. \nnb
\end{align}
We obtain the final result by using the probability generating functional of the Poisson point process \cite{chiu2013stochastic}. 
 	\end{IEEEproof}
 \begin{figure}
	\centering
	\includegraphics[width=0.9\linewidth]{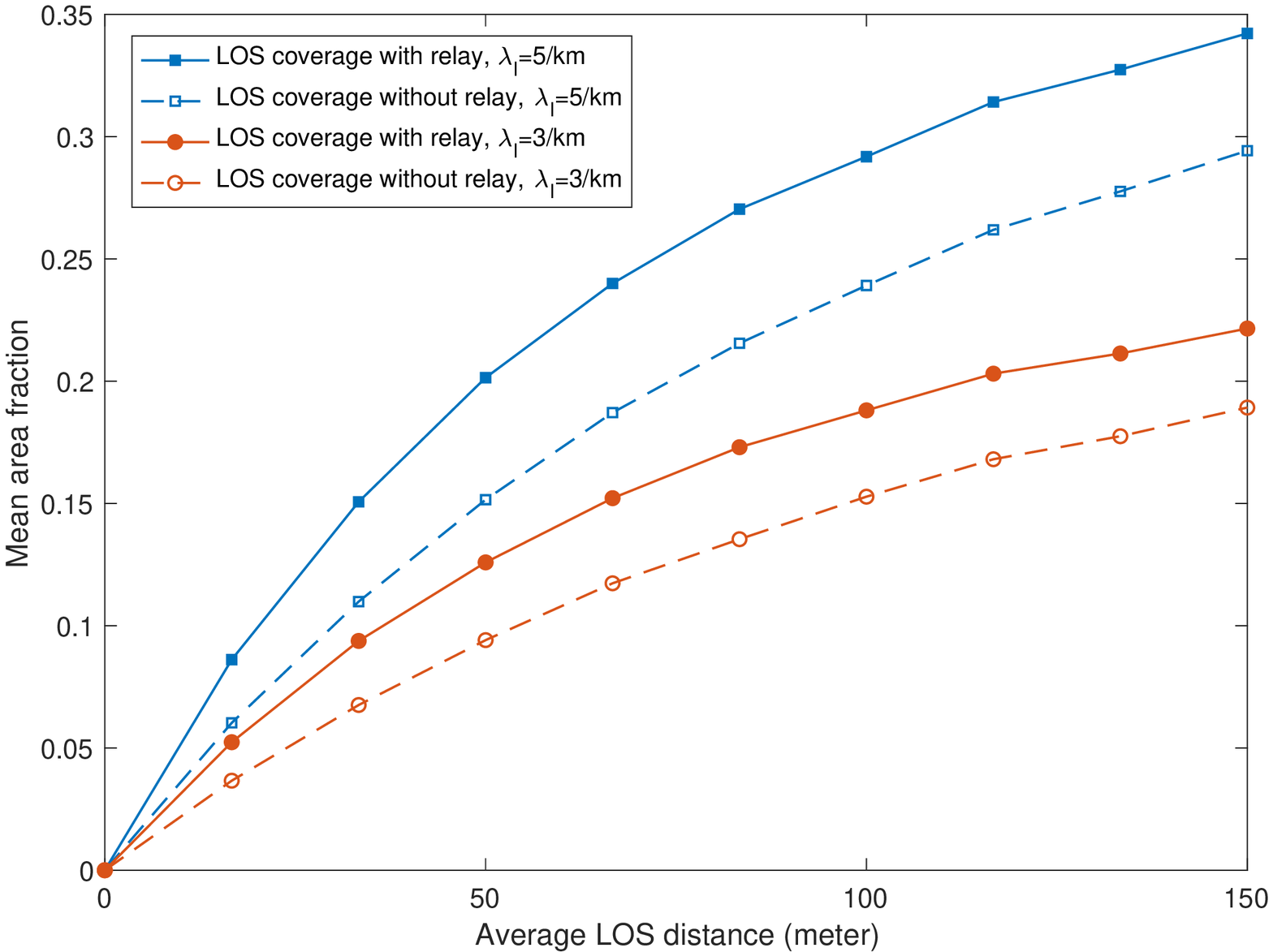}
	\caption{The mean area fractions of the LOS coverage of the proposed network where $ \mu=4 $/km and $ \eta=100 $ meters.}
	\label{fig:eta100lambda53mu4}
\end{figure}

\begin{figure}
	\centering
	\includegraphics[width=0.9\linewidth]{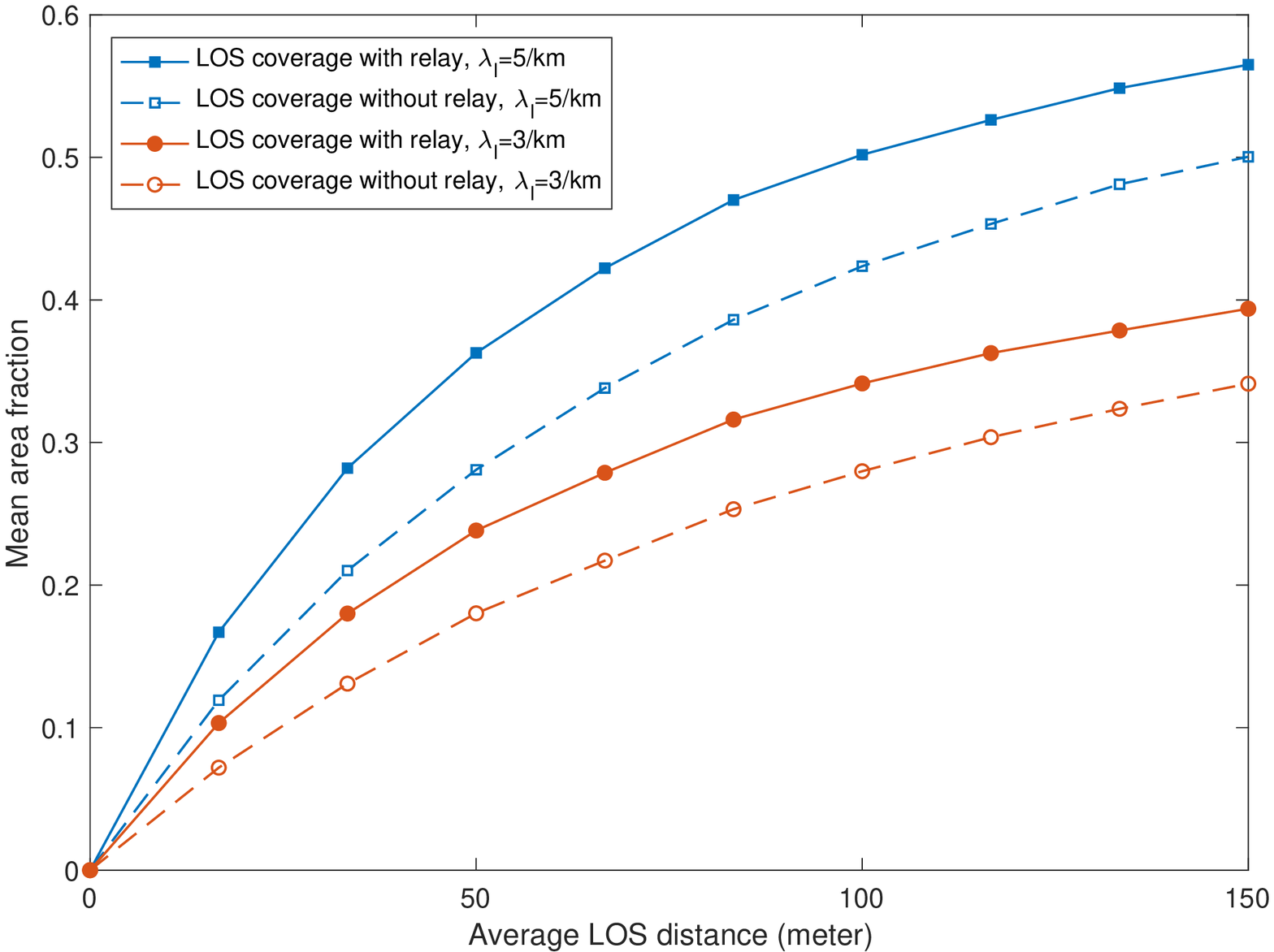}
	\caption{The mean area fractions of the LOS coverage of the proposed network where $ \mu=4 $/km and $ \eta=200 $ meters. } 
	\label{fig:rsuvsrelaycoverage}
\end{figure}
\begin{figure}
	\centering
	\includegraphics[width=0.9\linewidth]{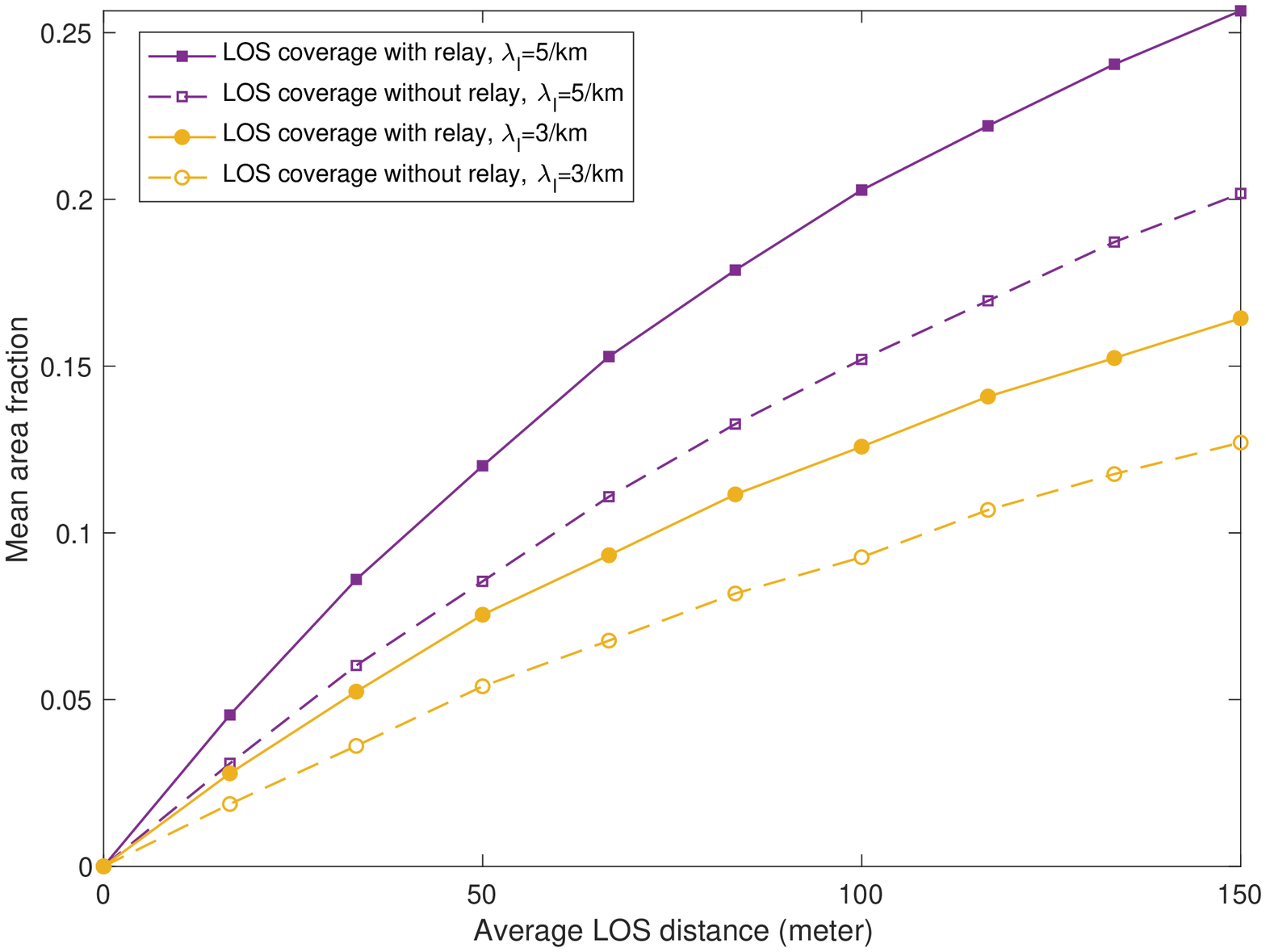}
	\caption{The mean area fractions of the LOS coverage of the proposed network where $ \mu=2 $/km and $ \eta=100 $ meters.}
	\label{fig:eta100lambda53mu2}
\end{figure}
\begin{figure}
	\centering
	\includegraphics[width=0.9\linewidth]{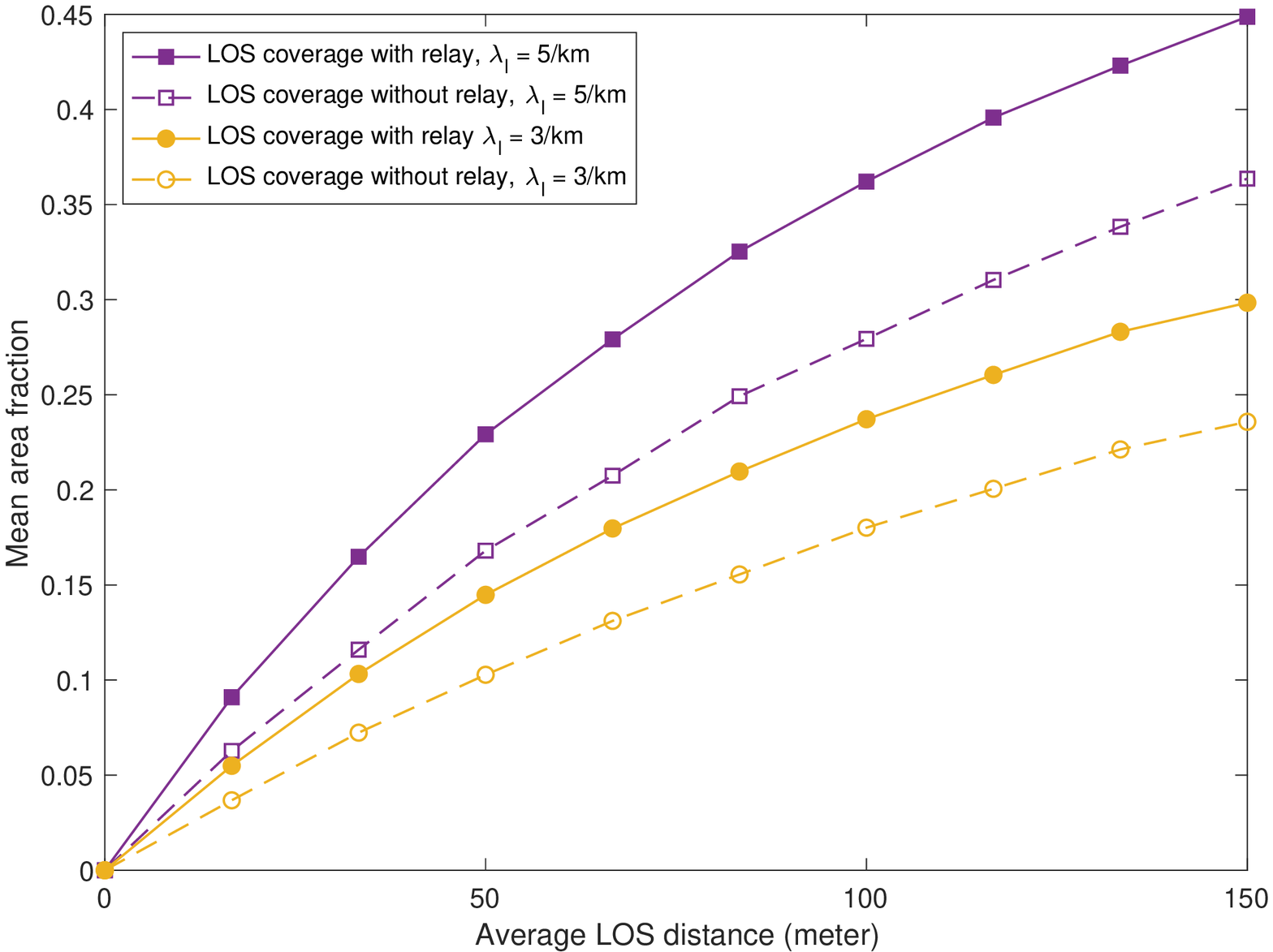}
	\caption{The mean area fractions of the LOS coverage of the proposed network where $ \mu=2 $/km and $ \eta=200 $ meters.}
	\label{fig:rsuvsrelaycoveragemu2}
\end{figure}
\begin{figure}
	\centering
	\includegraphics[width=0.9\linewidth]{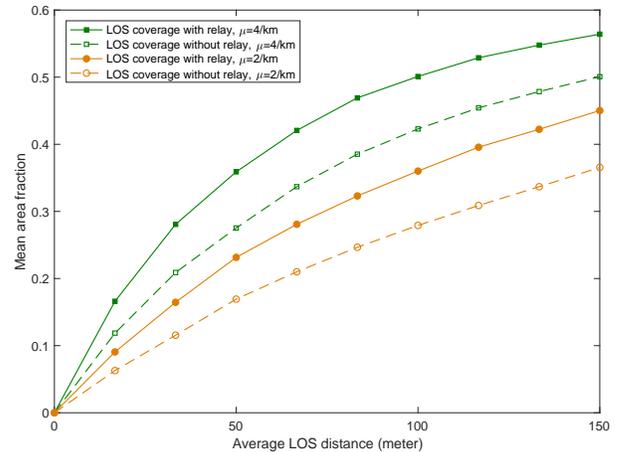}
	\caption{The mean area fractions of the LOS coverage of the proposed network where $ \lambda_l=10/\text{km}$ and $ \eta=100 $ meters. }
	\label{fig:eta100lambda10mu42}
\end{figure}
\begin{figure}
	\centering
	\includegraphics[width=0.9\linewidth]{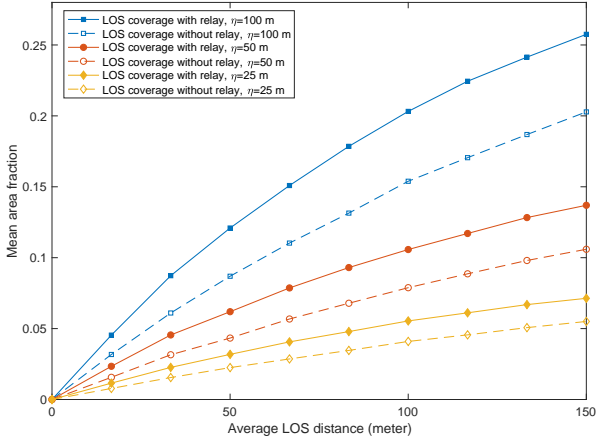}
	\caption{The mean area fractions of the LOS coverage of the proposed network where  $\lambda_l=5/\text{km} $ and $\mu=2/\text{km}.$ }
	\label{fig:variousetaslambda5mu2}
\end{figure}

Figs. \ref{fig:eta100lambda53mu4} and \ref{fig:rsuvsrelaycoverage} display the LOS coverage for $ \eta=100 $ and $ \eta=200 $ meters, respectively. We use $ \mu=4 $/km, i.e., the average inter-RSU distance of $250 $ meters. Figs. \ref{fig:eta100lambda53mu2}  and \ref{fig:rsuvsrelaycoveragemu2} show the LOS coverage for $ \eta=100 $ and $ \eta=200 $ meters, respectively where $ \mu=2 $/km, i.e., the average inter-RSU distance of $ 500  $ meters. Fig. \ref{fig:eta100lambda10mu42}  shows a vehicular network scenario where roads are densely distributed. Here, we use $ \lambda_l=10 $/km and $ \eta=100 $ meters.  In Fig. \ref{fig:variousetaslambda5mu2},  we analyze the increment of the LOS coverage as $ \gamma $ increases for $\eta = 25 $, $ 50, $ and $ 100 $ meters. When $ \gamma= 66 $ meters and $ \eta=25 $ meters, we have $ \nu(C_{\text{RSU}+\text{relay}})/ \nu(C_{\text{RSU}}) = 1.42. $ In other words, we have a nearly $ 40 $\% increase of the LOS coverage thanks to relays. Similarly, for $ \gamma=66 $ meters and $ \eta=50$ meters, $ \nu(C_{\text{RSU}+\text{relay}})/ \nu(C_{\text{RSU}}) = 1.39 $, i.e., the LOS coverage area increases by $ 39  $\%. For $ \gamma=66 $ meters and $ \eta=100 $ meters,   $ \nu(C_{\text{RSU}+\text{relay}})/ \nu(C_{\text{RSU}}) = 1.36 $, i.e., the LOS coverage increases $ 36 $\%. 
	
	We identify that the increments obtained from relays are greater for narrower roads.  In addition, based on Fig. \ref{fig:eta100lambda53mu4} through \ref{fig:variousetaslambda5mu2}, we show that the LOS gains are greater for smaller $ \gamma $s. For instance, in urban areas with large numbers of vehicles and obstacles, the increment of LOS coverage substantial.

\begin{remark}
	\cc{Our work assumes that the density of vehicles is much greater than the density of RSUs. This assumption ensures that each RSU can find at least one vehicle for relaying. Nevertheless, if the densities of RSUs and vehicles are similar, one should consider the following: (i) an RSU may not be able to find any potential vehicle within its LOS coverage and (ii) a vehicle can be selected by more than one RSU at the same time. Combined with the complications presented by the mobility of vehicles, these aspects introduce an additional layer of statistical dependence. The LOS coverage analysis for this case is left for future work. }
\end{remark}

\begin{remark}
\cc{The presented LOS coverage increment is applicable not only to the proposed random relay selection technique but also to other relay selection techniques. Since we randomly choose relays out of the LOS segments of RSUs, the increment of LOS coverage shown in this paper corresponds to the average of the increments of LOS coverage that would have been achieved by all of the relay candidates located within the LOS segments. As a result, the presented analysis of the LOS coverage gives a general insight that relays will expand the LOS coverage, enhance the chance that network users get LOS transmissions, and finally increase the service areas of LOS-critical applications in modern vehicular networks.}
\end{remark}

\begin{remark}
	\cc{A nonuniform relay selection technique may yield a higher area fraction for the LOS coverage. For instance, RSUs may select vehicles that are the furthest from RSUs to maximize the LOS coverage. In this case, each RSU is required to know the distances to all of its vehicles within its LOS coverage. In another example, a number of RSUs may jointly select their relays to globally maximize the LOS coverage. In this case, RSUs need to communicate not only with each other but also with relays. The analysis based on different relay selection techniques are left for future work. }
\end{remark}

\begin{remark}
	We assume that vehicles move along their roads with a constant speed. Nevertheless, the analysis presented in this paper is applicable to different mobility models as long as the  distribution of the vehicles is time-invariant \cite{8419219,8796442}. For instance, if each vehicle follows its own line and it chooses its own speed based on a given distribution, the locations of vehicles is a Poisson point process by displacement theorem \cite{baccelli2010stochastic}. Thus, the distribution of vehicles is time invariant. The vehicle point process is time-invariant and the analysis in this paper holds for that vehicle mobility model.
\end{remark}


\section{Discussions}

\subsection{Additive Approach to LOS Coverage}\label{S:4-A}

\cc{We have computed the mean area fractions of the RSU LOS coverage and RSU-plus-relay LOS coverage, by leveraging the stationarity of the particle model and then by computing the chance that the origin is contained by the particle model. If network users are uniformly distributed on roads, the mean fractions give the average number of LOS-accessible network users per unit space. }

\par On the other hand, one may consider an additive approach to evaluate the size of LOS coverage by ignoring the overlap of LOS coverage from RSUs on different roads. In this case, the mean area fraction of the RSU LOS coverage is  
\begin{align}
	\nu(C_{\text{RSU}}) \approxeq \lim_{r\to\infty}\dfrac{\eta \times \text{length}\left(B_0(r)\cap \left(\bigcup_{X_i\in\Phi}S_{X_i}\right)\right)}{\text{area}(B_0(r))}. 
\end{align}
To evaluate the right-hand side, we use the fact that the total length of the Poisson lines in a disk of radius $ r $ is 
$ \pi\lambda_l r^2 $\cite{chiu2013stochastic}. Thus, we have  
\begin{align}
	&\eta \times \text{length}\left(B_0(r)\cap \left(\bigcup_{X_i\in\Phi}S_{X_i}\right)\right) \nnb\\
	&= \pi\eta  \lambda_l r^2 \nu_1\left(\cup_{X_i\in\phi}S_{X_i}\right)\nnb,
\end{align}
where $ \nu_1\left(\cup_{X_i\in\phi}S_{X_i}\right) $ denotes the \emph{linear} fraction of the LOS segments w.r.t. a line and $ \phi $ is the RSU Poisson point process of intensity $ \mu $ on that line. Since  the RSU point process is a stationary, the linear fraction of LOS segments is equal to the probability that a typical point is contained by the LOS segments of RSUs.  Therefore, the linear fraction is given by 
\begin{align}
	\nu_1\left(\cup_{X_i\in\phi}S_{X_i}\right)&=1-\bP\left(0\notin \bigcup\limits_{X_i\in\phi} S_{X_i}\right)\nnb\\
	&=1-\bE\left[\prod_{{X}_i\in\phi}\bE\left[\ind_{0\notin S_{X_i}}|\phi\right]\right]\nnb\\
 &=1- \exp\left(-2\mu\int_0^\infty \exp\left(-\frac{x}{\gamma}\diff x \right)\right)\nnb\\
	&=1-\exp\left(-2\mu\gamma\right).
\end{align}
As a result, based on the additive approach, the size of RSU LOS coverage is 
\begin{align}
	\nu(C_{\text{RSU}}) &\approxeq \lim_{r\to\infty}\dfrac{\eta \times\pi\lambda_lr^2 (1-\exp(-2\mu\gamma)) }{\pi r^2} \nnb\\
	&=\eta\lambda_l(1-\exp(-2\mu\gamma)).
\end{align}
\subsection{Error of Additive Approach}
\begin{figure}
		\centering
		\includegraphics[width=0.9\linewidth]{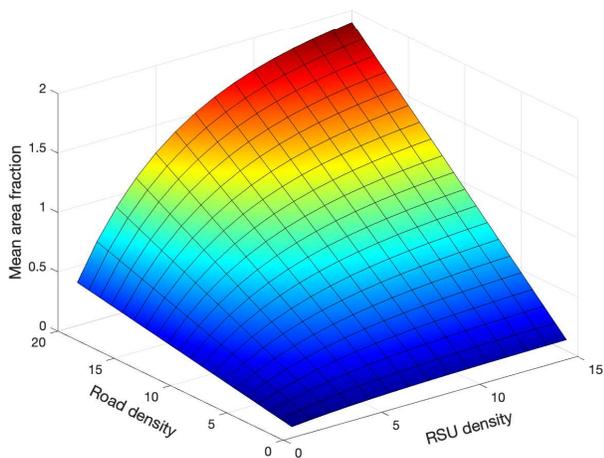}
		\caption{The difference between the exact analysis and 1-D approximation where $ \gamma = 100 $ meters and $ \eta =100$ meters. The units of road density and RSU density are per meter.}
		\label{fig:gap2d1deta100}
	\end{figure}

\begin{figure}
	\centering
	\includegraphics[width=0.9\linewidth]{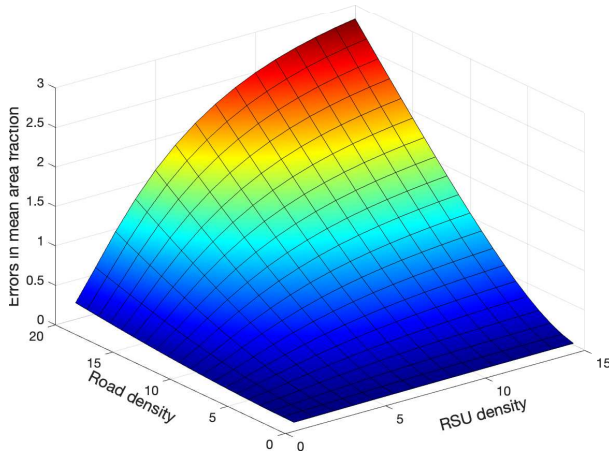}
	\caption{The difference between the exact analysis and 1-D approximation where $ \gamma = 100 $ meters and $ \eta =200$ meters. The units of road density and RSU density are per meter.}
	\label{fig:gap}
\end{figure}

In this section, we compare the RSU LOS coverage computed by the proposed framework and by the additive approach discussed in Section \ref{S:4-A}. The error $ \Gamma $ is given by 
	\begin{align}
		\Gamma= |(1-e^{-2\lambda_l\eta(1-\exp(-2\mu\gamma))}) - (\lambda_l\eta(1-e^{-2\mu\gamma})) |. 
	\end{align}
 Figs. \ref{fig:gap2d1deta100} and \ref{fig:gap} show the error between approximation approach and the exact analysis. Since the additive approach does not take into account for the overlap of LOS coverage from different road RSUs, the error increases as as $ \gamma $, $ \lambda_l $ or $ \mu $ increases.
 \par \cc{Based on this observation, we conclude that the overlap of LOS coverage should not be ignored and additive approach yields higher error when the spatial correlation of RSUs and relays is stronger. }

\section{Conclusion}
The availability of LOS signals dictates the success of modern V2X applications. In this paper, we investigate the potential of employing vehicle relays as a means to increase the LOS coverage of network users. We provide an analytical framework to model spatially correlated RSUs, vehicles, and their LOS coverage. By assuming that vehicle relays provide the LOS coverage only when they are in the LOS coverage of RSUs, we derive the mean area fraction of the RSU-plus-relay LOS coverage and show that relays increase the LOS coverage of the network and expand the service areas of LOS-critical applications. We show that relays can increase the area fraction of LOS coverage by nearly 50\% even when RSUs and relays are spatially correlated. The present analysis of the LOS coverage increment not only displays the benefit of employing relays but also assesses the viability of the modern V2X applications based on the LOS coverage.

 	\section*{Acknowledgement}
The work of Chang-sik Choi was supported in part by the NRF-2021R1F1A1059666 and by the Institute of Information and Communication Technology Planning and Evaluation (IITP) grant funded by the Korea Government (MSIT), QoE improvement of open Wi-Fi on public transportation for the reduction of communication expense, under Grant 2018-0-00792.  The work of Francois Baccelli was supported in part by the Simons Foundation grant 197982 and by the ERC NEMO grant 788851 to INRIA.
	\bibliographystyle{IEEEtran}
	\bibliography{ref}
\end{document}